\documentclass[12pt]{JHEP3}
\newcommand{\refeq}[1]{(\ref{#1})}
\usepackage{graphicx}
 \usepackage{ae}
 \usepackage{aecompl}
\usepackage{amsmath,epsfig}
\usepackage{amssymb,amsfonts}
\usepackage{latexsym}
\usepackage{epsfig}
\usepackage{subfigure}

\relax
\renewcommand{\theequation}{\arabic{section}.\arabic{equation}}

\def\be{\begin{equation}}
\def\ee{\end{equation}}

\newcommand{\<}{\langle}
\renewcommand{\>}{\rangle}
\newcommand{\de}{\partial}
\newcommand{\bear}{\begin{eqnarray}}
\newcommand{\bea}{\begin{eqnarray}}
\newcommand{\eear}{\end{eqnarray}}
\newcommand{\eea}{\end{eqnarray}}
\newbox\pippobox

\def\II{\relax{\rm I\kern-.18em I}}

\def\cO{{\cal O}}

\def\m{\mu}
\def\n{\nu}

\def\A{{\cal A}}
\def\O{{\mathcal O}}
\title{On the gluon operator effective potential in holographic Yang-Mills theory}

\author{Elias Kiritsis$^{1,2}$, Wenliang Li$^2$ and Francesco Nitti$^2$\\
 ~\\
 $^1$
 \href{http://hep.physics.uoc.gr/}
 {Crete Center for Theoretical Physics}, Department of Physics, University of Crete
 71003 Heraklion, Greece\\
  ~\\
$^2$
\href{http://www.apc.univ-paris7.fr}
{APC, Universit\'e Paris 7}, CNRS/IN2P3, CEA/IRFU, Obs. de Paris, Sorbonne Paris Cit\'e, B\^atiment Condorcet, F-75205, Paris Cedex 13, France (UMR du CNRS 7164).}

\abstract{The holographic formalism is applied to the calculation of the effective potential for the scalar glueball operator. Three different versions of this operator are defined, and for each we compute the associated effective potential and discuss its properties and scheme ambiguities. Contact is made to earlier attempts to guess this effective potential from the conformal anomaly. We apply our results to the Improved Holographic QCD model calculating the glueball condensate.}


\preprint{CCTP-2014-19\\ CCQCN-2014-43}

\begin{document}

\maketitle 


\section{Introduction} \label{intro}

The $AdS$/CFT correspondence can  potentially address many non-perturbative features of four-dimensional QCD-like  gauge theories. As it has become clear in the past 15 years, virtually all the observables one can construct in QCD can be associated to  a geometric counterpart in holographic models.

Although string theoretical, top-down constructions only allow today precise first-principle calculations in supersymmetric versions of Yang-Mills/QCD and deformations thereof, a parallel effort has been undertaken in order to construct, using the same holographic dictionary, phenomenological models which may be closer to QCD \cite{gkn,gubser,thermo2,thermo3,ihqcd}, using bulk Einstein gravity coupled to a scalar field and considering full backreacted
solutions of the system.

In the same  context, it was recently  shown in  \cite{kln} how to write a    fully non-linear expression for the generating functional of scalar external sources up to two space-time derivatives  and, by  a Legendre transformation, for the quantum one-particle irreducible effective action for the associated classical operator.

In this work we provide an application of the general results of \cite{kln} in the context of QCD.   We concentrate on one particular observable, which is present in any non-abelian gauge theory: the  gauge-invariant, scalar, dimension four,  gluonic operator:

\be\label{G}
{\cal G} = Tr F^{\mu\nu}F_{\mu\nu}.
\ee

The vacuum expectation value $\<{\cal G}\>$ has been widely discussed in the past. It is one of the main ingredients of the SVZ sum rules \cite{svz}, which relate expectation values of QCD composite operators to hadronic observables. Various numerical studies have  been performed to compute this quantity using lattice gauge theory \cite{lattice}\footnote{Some of those works focused on the operator (\ref{G}), others on its renormalization group-invariant version $(\beta(g)/g^3)Tr F^2$, where $\beta(g)$ is the beta function of the theory. At one loop order, these two operators coincide.   In this paper we will discuss both versions of the dimension-four gluonic operator.}.
In pure Yang-Mills theory in the continuum limit, the expectation value   diverges as $a^{-4}$, where $a$ is the lattice spacing, and it is also subject to  an additive renormalization. Other works concentrated on  the temperature dependence  of ${\cal G}$ in the deconfined phase which, contrary to the vacuum expectation value, is unambiguous, and  once the vacuum contribution is subtracted, it is related  to the interaction measure (i.e. the trace of the thermal stress tensor) \cite{latticeT}.

The quartic divergence of the one-point function $\<{\cal G}\>$ is also present in the holographic theory and it arises close to the UV boundary of the asymptotically $AdS$ space-time.
Renormalization of this quantity proceeds by the same counterterm that renormalizes the (also quartically divergent) vacuum energy. However, the finite part which is left over is scheme-dependent, and subject to an additive renormalization. Therefore it is not very interesting to compare the (finite) result  $\<{\cal G}^{(ren)}\>$  obtained after renormalization to the numerical values  one can find in the literature (from lattice computations, or from phenomenology coupled to sum rules), unless one can relate it to other physical quantities or establish a precise map  between renormalization schemes.

The $AdS$/CFT duality  however allows the possibility to go   beyond  the calculation of the  (scheme-dependent) value of the condensate, and to compute the  {\em full quantum effective potential} for a QFT  operator  ${\cal O}$. This is the homogeneous part of the one-particle-irreducible generating functional,
\be\label{gamma}
\Gamma({\cal O}) = \int d^4 x\, J\, {\cal O} \,\,- \,\,  S,
\ee
where $J$ is the source coupling to the operator ${\cal O}$ and $S$ is the renormalized generating functional for connected correlators, computed by the renormalized gravity on-shell action\footnote{Throughout the paper, we work in  the Euclidean signature, in which equation (\ref{gamma}) is the appropriate definition of the quantum effective action. See the note about conventions at the end of the introduction.}. The zero-derivative part of  functional $\Gamma({\cal O})$, i.e. the quantum effective potential for the field ${\cal O}$,  is uniquely determined once the vacuum energy is renormalized  to a definite value. 
Further, the computation can be extended to include higher derivative terms. 

In this paper we present  the calculation of the  quantum effective action for the  gluonic operator (\ref{G}) and for its RG-invariant relative, in phenomenological holographic duals of pure Yang-Mills theory constructed from Einstein-Dilaton five-dimensional models, and in particular in the Improved Holographic QCD model \cite{ihqcd}. These models are characterized by the presence of a bulk scalar field, which is dual to the  source associated to this  operator i.e. the Yang-Mills coupling $\lambda$. The dual scalar field  has a non-trivial dependence on the radial coordinate, which encodes the RG-scale dependence of the coupling.\footnote{Our setup and calculation can be easily extended to the multiscalar case, encoding the potential of multiple scalar operators.}

We will compute the effective action up to two space-time derivatives, i.e. in the general form: 
\be\label{intro1}
\Gamma[{\cal O}] = \int d^4 x\, \left[ {\cal V}({\cal O}) + {1\over 2} G({\cal O}) \de_\mu {\cal O}\de^\mu {\cal O}\right], 
\ee
and we will compute the effective potential ${\cal V}({\cal O})$ and the kinetic function $G({\cal O})$.  We will also present the results for the corresponding canonically normalized field, obtained by integrating the relation 
$d{\cal O}^{c} = G({\cal O})\, d {\cal O}$. 

In the holographic theory, we  show that one can define a non-perturbative, RG-invariant scale $\Lambda$ \cite{ihqcd}, which is the geometric analog of the QCD IR scale, and sets the scale of all non-perturbative observables (condensates, glueball masses, etc).   We compute the effective potentials both in the bare theory with a
UV cut-off $\Lambda_{UV}$, and in the renormalized theory at a scale $\mu$.
 The  effective potentials  depend on the scale $\Lambda$ and, depending whether we are using the renormalized or the bare regularized theory, on the RG scale $\mu$ or cut-off scale $\Lambda_{UV}$, respectively.

In the renormalized theory, we consider the effective potentials corresponding to different but related operators:
\begin{enumerate}
\item The first one is the gluonic   operator defined in equation (\ref{G}). The source for this operator in the field theory is the 't Hooft coupling, which is scale-dependent. Thus, the renormalized effective potential will depend on the renormalization scale $\mu$, and has the general form:
\be\label{gammaren}
{\cal V}[{\cal G}, \mu] = \Lambda^4 \, v\left({{\cal G}\ \over \Lambda^4}, {\mu \over \Lambda} \right).
\ee
where $v$ is a dimensionless function of its arguments, whose precise form depends on the bulk dilaton potential\footnote{More precisely, it depends on the superpotential associated to the bulk solution.}, and $\mu$ is the renormalization scale.  The value of the non-perturbative scale $\Lambda$ is fixed by choosing a particular solution of the RG-flow equations, which in the bulk amounts to fixing all integration constants of Einstein's equations.

Beside being scale-dependent, the effective potential  for $Tr F^2$  depends crucially on the relation between the field theory 't Hooft coupling $\lambda$ and the bulk scalar field. This is one of the sources of ambiguity  of  phenomenological models, as it is not fixed from general principles. In the UV, this ambiguity can be fixed to a certain extent, by matching the holographic model to Yang-Mills perturbation theory. However, changing how we identify the coupling in the IR can be interpreted as a source of scheme dependence, and it  modifies the shape of the potential.
The precise identification of the coupling   can only be fixed if the gravity dual is embedded in a top-down model.

Thus, although the calculation of $\Gamma({\cal G})$ can be performed, it is a nontrivial issue to relate the result to field theory non-perturbative observables,   without a clear  identification of the coupling with a bulk field at all scales. This drawback is avoided if we consider the renormalization-group invariant gluonic operator, as we discuss next.

\item The renormalization group invariant version of the
gluonic operator $\cal G$  is defined by:
\be\label{T}
{\cal T} = -{\beta(\lambda) \over 2\lambda^2} Tr F^2,
\ee
where $\beta(\lambda)$ is the beta-function for the running 't Hooft's coupling  $\lambda$.
This operator is related by the conformal anomaly to the trace of the stress tensor, ${T^\mu}_\mu = {\cal T}$.
Unlike the case described in point 1 above, the form of the effective potential for ${\cal T}$ is universal, it does not depend on the identification of the coupling, nor on the precise form of the bulk dilaton potential,  and it can be written in closed analytic form:
\be \label{gammarenT}
{\cal V}[{\cal T}] = {{\cal T} \over 4} \left(\log {{\cal T} \over \Lambda^4} - 1 \right),
\ee
where $\Lambda$ is the holographic non-perturbative scale.
Due to the RG-invariance of the operator ${\cal T}$, the potential is itself independent of scale. Its expression is universal up to the scheme-dependent vacuum energy renormalization, which in writing equation (\ref{gammarenT}) was chosen in such a way that ${\cal V}({\cal T})$ is extremized exactly at the
non-perturbative scale $\Lambda$:
\be
{\de {\cal V} \over \de {\cal T}}\Big|_{{\cal T} = \Lambda} = 0.
\ee

Equation (\ref{gammarenT}) reproduces  the field theory trace anomaly, and in a sense it is an exact, model-indepednent prediction  of holography applied to QCD.

In the UV, at small coupling  $\beta(\lambda) \sim -b_0\lambda^2$, and the operators (\ref{G}) and (\ref{T}) coincide up to a multiplicative constant. In fact, a UV expansion of our computation shows that,  as $\mu \gg \Lambda$, the effective potential (\ref{gammaren}) reduces to the result (\ref{gammarenT}).

\item Going  beyond the effective potential, we consider the two-derivative term in (\ref{intro1}), which can be obtained from the general results in  \cite{kln}, and   derive effective actions for the {\em canonically normalized} operators corresponding to ${\cal G}$ and ${\cal T}$. The corresponding  effective action up to two derivatives (and omitting the Einstein-Hilbert term) has the form:
\be\label{gamma2}
\Gamma[{\cal O}^{(c)}] = \int d^4 x \, \left[\Lambda^4 { v} \left({{\cal O}^{(c)} \over \Lambda}, {\mu\over \Lambda }\right) + {1\over 2} \de_\mu {\cal O}^{(c)} \de^\mu {\cal O}^{(c)}\right].
\ee
where ${\cal O}^{(c)}$ stands for either  ${\cal G}^{(c)}$ or ${\cal T}^{(c)}$. In both cases, the form of the potential  ${\cal V}$  in this case is not universal, but depends on  the details of the bulk theory.
\end{enumerate}

For the operators discussed above, we also compute  the bare effective action,  calculated  in the theory regularized  at a cut-off scale $\Lambda_{UV}$. This is a  useful quantity to compare with lattice calculations before renormalization: it is independent of the renormalization procedure, which may be difficult to translate between
holography and other techniques. The quartically divergent effective potential in the bare theory takes the general form:
\be\label{gammareg}
{\cal V}^{(reg)}[{\cal O},\Lambda_{UV}] = \Lambda_{UV}^4  \,\, v^{(reg)}\left({{\cal O}\ \over \Lambda_{UV}^4}, {\Lambda \over \Lambda_{UV}} \right) ,
\ee
where ${\cal O}$ stands for either ${\cal G}$ or ${\cal T}$. In the equation above,  $\Lambda_{UV}$ sets the cut-off scale, whereas $\Lambda$ is set by the choice of the initial condition (the value of the source/coupling) at the cutoff.  The  cut-off is a priori arbitrary, but for the theory to be perturbative in the UV (as in standard QCD) one must require that $\Lambda/\Lambda_{UV} \ll 1$. In order to make more explicit the connection with field theoretical computations, we also provide the explicit form of the subtraction of the vacuum energy, as a function of $\Lambda_{UV}$ and $\Lambda$, necessary to go from the regularized to the renormalized theory.

Except for the exact RG-invariant effective action (\ref{gammarenT}), the full effective potentials we discuss can only be found numerically, once a concrete bulk theory is specified.  In the final part of the paper we provide explicit numerical results for the effective potentials (\ref{gammaren}) and (\ref{gammareg}), and compare them with the universal function  (\ref{gammarenT}),  in a concrete theory:  the Improved Holographic QCD model (IHQCD), reviewed in  \cite{ihqcd},  which was shown  to lead to realistic, quantitative agreement  with lattice Yang-Mills  spectra and thermodynamics, \cite{thermo3}.

There are several  aspects of our computation  that are of interest, both for QCD application and more generally for the phenomenology of strongly interacting field theories. On the QCD side, although we are not aware of a lattice calculation of the effective potential of $Tr F^2$, such a calculation should in principle be possible, and it would be interesting to compare the result to the one found here after the ambiguiity in the normalization of the extremum  of the potential are removed.

It has been proposed that a strongly coupled sector may account for the dynamics observed in nature (for example for electroweak symmetry breaking, or for inflation \cite{baumann}), and holography has been often used as a model-building tool to describe the strongly coupled sector (see e.g. \cite{piai} for a recent investigation of holography in the context of technicolor-like theories).
Our results for the effective action for the canonically normalized operator, including the kinetic term, can serve as the starting point for phenomenological studies in theories where the vacuum dynamics is not driven by an elementary field, nor  by (composite) particle-like excitations, but rather by the dynamics of a condensate. In fact, the effective action (\ref{gamma2}) should not be interpreted as  describing particles arising from the quantization of ${\cal O}$, and it is not to be confused with e.g. the effective actions of particle {\em states} (glueball excitations) in a strongly coupled theory. It is therefore quite different from those described
in e.g. \cite{megias}, where the attention was focused on particle-like excitations in holographic theories, and the computation of properties such as their mass.

The structure of this work is as follows.
In Section 2 we review the holographic five-dimensional Einstein-dilaton setup used to describe Yang-Mills theory,  we provide  the holographic dictionary, and we give the holographic definition of the  non-perturbative scale $\Lambda$.

In Section 3 we define the various operators we consider and provide  the general procedure to construct the (renormalized or regularized) quantum effective actions. We also give the explicit form of the subtraction procedure used to renormalize the vacuum energy.

In Section 4 we give the results for the effective potentials, computed numerically in the concrete IHQCD background.

In Section 5 we offer some concluding remarks and perspectives.

Technical details of the UV and IR form of the effective potential for $Tr F^2$ are left for the Appendix.

\subsection*{Notation and conventions} Throughout the paper we will work in the Euclidean signature, in which the appropriate definition of the Legendre transofrm is given by 
\be
\Gamma  = \int  J\, {\cal O} \,\,-\,\, S\,,
\ee
 and the relation between sources and  one-point functions is: 
\be
{\cal O} = {\delta S[J] \over \delta J}, \qquad J ={ \delta \Gamma[{\cal O}] \over \delta {\cal O}}
\ee
(no extra  $i$'s needed, as it is the case instead in the Minkowski signature). Given the Euclidean effective action in the form (\ref{intro1}), the corresponding   Minkowski effective action is readily written as:
\be \label{Gamma-mink}
\Gamma_M[{\cal O}] = \int d^4 x\, \left[ -{\cal V}({\cal O}) - {1\over 2} G({\cal O}) \eta^{\mu\nu}\de_\mu {\cal O}\de_\nu {\cal O}\right]. 
\ee
with $\eta_{\mu\nu} = diag(-1,+1,+1,+1)$ and  the same functions ${\cal V}({\cal O}) $ and $G({\cal O})$ as in (\ref{intro1}). 

We denote boundary space-time indices by $\mu,\nu\ldots$, bulk space-time  indices by $a,b,\ldots$,  we use $g_{ab}$ for the bulk metric and we keep the notation $\eta_{\mu\nu}$ for the flat euclidean metric.

\section{Yang-Mills theory and Einstein-dilaton gravity}

\subsection{The Model}

In holography, pure 4d Yang-Mills theory at large $N_c$ is expected to be dual to a 5d string theory (or a classical gravitational theory in the low energy limit). There is one extra dimension in the dual theory because of the existence of a single adjoint vector  in the boundary YM theory \cite{diss,ihqcd}.

The gauge invariant single-trace operators and the bulk on-shell states are in one-to-one correspondence.
The most relevant degrees of freedom are the lowest dimension gauge-invariant operators $Tr[F_{\mu\nu}F_{\rho\sigma}]$, which can be decomposed according to spin into: the stress tensor $T_{\mu\nu}\sim Tr[F_{\mu\rho}{F^\rho}_\nu-\frac 1 d \eta_{\mu\nu}F^2]$, the YM operator $Tr[F^2]$ and the topological density $Tr[F\wedge F]$. In the holographic dual, the related bulk fields are the 5d metric $g_{\mu\nu}$ (dual to the stress tensor $T_{\mu\nu}$),  a scalar field $\phi$ (dual to the YM operator $Tr[F^2]$) and an axion field dual to  $Tr[F\wedge F]$. In the following, we will ignore the axion because the contribution to the QCD vacuum from the topological sector  is suppressed by  $\frac 1 {N_c}$ in the large $N_c$ limit. 

 To be specific, we consider a bulk theory with the following five-dimensional action:
 \begin{equation}
S=S_{bulk}+S_{GH},
\end{equation}
where $S_{bulk}$ are the bulk terms:
\begin{equation} \label{action}
S_{bulk}= -M^{3}\int \,d^4 x \,d u\sqrt{-g}\left[R^{(5)}-\frac 1 2g^{ab}\partial_a\phi\partial_b\phi+V(\phi)\right],\end{equation}
and $S_{GH}$ is the boundary Gibbons-Hawking (GH) term:
\begin{equation}
S_{GH}= 2M^{3}  \int \,  \left. d^4 x \big(\sqrt {-\gamma} K\big) \right|_{UV}^{IR},
\end{equation}
In the above expressions, the coordinate $u$ is the radial (or holographic) coordinate, and   $\{x^\mu \}_{\mu=0\ldots 3}$ are the space-time coordinates on constant-$u$ hypersurfaces. $M$ is the five-dimensional Planck scale.  The number of colors $N_c$ sets the magnitude of the five-dimensional Planck scale: in the large $N_c$ limit we assume that
\be
M^3\propto N_c^2, 
\ee
while  the scalar potential $V(\phi)$ stays of order one,  so that classical gravity can be used as a reliable approximation.

The bulk scalar potential  encodes all the information about the RG flow.
The holographic direction, which is related to the energy scale of boundary field theory, is parametrized by the $u$-coordinate.
The ultraviolet and the infrared endpoints of this coordinate are denoted by $u_{UV}$ and $u_{IR}$,
which may be the physical  $UV$ and $IR$ fixed points of the full theory, or the $UV$ and $IR$ cutoffs.
In the Gibbons-Hawking term,  $\gamma_{\mu\nu}$ is the induced
metric on the  slices and $K=\gamma^{\mu\nu}K_{\mu\nu}$ is the trace
of the extrinsic curvature.  Here and in the following discussions, the subscripts $UV$ or $IR$ indicate  that the quantities are evaluated on the $UV$ or $IR$ slices.

Vacuum solutions which preserve Poincar\'e invariance have the general domain-wall form (up to diffeomorphisms):
\be\label{metric}
ds^2 = du^2 + e^{2A(u)}\eta_{\mu\nu}dx^\mu dx^\nu, \qquad \phi=\phi(u).
\ee

Solutions of this form may be  found by first specifying  a superpotential $W(\phi)$, i.e. a particular solution of the  equation:  
\be
V=\frac 1 3 W^2-\frac 1 2{W^\prime}^2,\label{superpotential-eq}
\ee
where $'=\frac {d}{d\phi}$ is the derivative with respect to the scalar field.

Once $W$ is chosen\footnote{The superpotential should be  chosen in such a way as to satisfy appropriate IR regularity conditions. For example it should be such that the solution allows small black hole deformations \cite{gubser1}, and/or that the fluctuation problem is well-posed without the need of extra IR boundary conditions \cite{5dgraviton}. For monotonic potentials these criteria  specify the solution uniquely \cite{thermo2}.}, the functions $(A(u),\phi(u))$ are solutions of the first order system:
\be\label{sol1}
{dA\over du} = -{1\over 6}W(\phi), \qquad {d\phi \over du} = W'(\phi).
\ee
The solution (together with a specified initial condition\footnote{Due to the shift symmetry in $u$ of the system (\ref{sol1}) one can impose this initial condition at an arbitrary $u$ without physically changing the solution, i.e. different choices of the initial point can be related by a diffeomorphism. What is invariant is the value of $A$ at a given $\phi$.}) may be put  in the form:
\be\label{sol2}
A(u) = {A}^\ast - {1\over 6}\int^{\phi(u)}_{\phi^\ast} d\phi {W \over W'}, \qquad \left\{\begin{array}{l} A(u^\ast) ={A}^\ast \\ \phi(u^\ast) = \phi^\ast \end{array} \right.\,.
\ee
We will consider solutions that have a UV-asymptotically AdS region, where 
\be
A(u) \sim -u/\ell \qquad \text{as} \qquad u \to \infty.  
\ee
This equation sets the asymptotic AdS scale, which appears in the asymptotic value of the potential (and of the superpotential):
\be
\lim_{u\to -\infty} V(\phi(u)) = {12 \over \ell^2},  \qquad \lim_{u\to -\infty} W(\phi(u)) = {6 \over \ell}.
\ee 
\subsection{The holographic dictionary and the  potential}

Given a solution of the form (\ref{metric}) , the holographic dictionary  between gravity and field theory quantities is the following:

\begin{enumerate}
\item The field theory energy scale $\mu$ is identified with the metric scale factor $e^A$, up to a multiplicative constant, which we can choose to be
the asymptotic $AdS$ scale:
\be\label{mu}
\mu = \ell^{-1} e^A.
\ee
This is the appropriate (and reparametrization-invariant) translation between the radial coordinate and the energy scale \cite{kln}.
\item
We define the field $\lambda(r)$ as:
\begin{equation}\label{lambda}
\lambda=e^{\sqrt{\frac 3 8}\phi}.
\end{equation}
As discussed in  \cite{ihqcd}, this field is identified with the running  't Hooft coupling of the QFT\footnote{In \cite{ihqcd} the scalar field is not canonically normalized and the 't Hooft coupling is $\lambda=e^\phi$. Here the scalar field $\phi$ is canonically normalized, which accounts for the additional factor $\sqrt {\frac 3 8}$ in the definition of 't Hooft coupling.}. As $\lambda\rightarrow 0$, i.e. in the UV limit,  the identification with 't Hooft coupling is accurate. However, as discussed in detail in \cite{thermo3}, this assumption may not hold in the IR, and an extra field redefinition may be necessary to relate the bulk field $\phi(r)$ to the gauge coupling. This introduces an ambiguity in the calculation
of IR quantities, like the quantum effective potential for $Tr F^2$, which relies on a particular identification of the 't hooft coupling with a particular bulk quantity. 

However, as we will see in section \ref{invariant}
the effective action for an appropriate RG-invariant gluon condensate will be independent of the identification of 't Hooft coupling with a specific function of the bulk field $\phi$.
\item The holographic $\beta$-function is obtained from (\ref{sol1}) and (\ref{mu}-\ref{lambda}):
\begin{equation} \label{beta}
\beta(\lambda)\equiv\frac{d\lambda}{d\log \mu}=\frac{d\phi}{d A}\frac{d\lambda}{d\phi}=- {9\over 4} {\lambda^2 \over W} {\de W \over \de \lambda} .
\end{equation}
\end{enumerate}
Using equations (\ref{sol2}), (\ref{mu})  and (\ref{beta}) we recover the usual integrated RG-flow relation between the scale and the running coupling,
\be\label{beta2}
\log \mu = \log{{\mu}^\ast} + \int_{{\lambda}^\ast}^{\lambda(\mu)} {d\lambda \over \beta(\lambda)}.
\ee

In the following, we will review the choice of the bulk scalar potential for realistic holographic models like IHQCD \cite{ihqcd}.

Let us start with the small-$\lambda$ behavior of the potential. In the UV, an asymptotically free theory is also asymptotically scale invariant. Thus the corresponding bulk geometry should tend towards $AdS_5$ spacetime, and the bulk potential should asymptote to the cosmological constant $12/\ell^2$. 
Then the coefficients of the UV expansion around the constant are fixed by the perturbative $\beta$-function through equations (\ref{beta}) and (\ref{superpotential-eq}).  

Concretely, in Yang-Mills theory the 1-loop $\beta$-function is
\begin{equation}\label{betaqcd}
\beta_{QCD}=-b_0\lambda^2+O(\lambda^3).
\end{equation}
where $b_0=\frac {11}{24\pi^2} $.
The first two leading terms in the UV expansion of the superpotential $W(\lambda)$ are therefore fixed by  matching the holographic $\beta$-function  (\ref{beta}) and YM $\beta$-function (\ref{betaqcd}):
\begin{equation}
W=\frac 6 \ell \left(1+\frac {11}{54\pi^2} \lambda+ O(\lambda^2)\right),\label{W-UV}
\end{equation}
The UV-asymptotic form of the bulk scalar potential is determined by the above  equation  via equation (\ref{superpotential-eq}): 
\begin{equation}\label{V-UV}
V=\frac {12}{\ell^2}\left(1+\frac {11}{27\pi^2}\lambda+O(\lambda^2)\right).
\end{equation}

At large $\lambda$, we consider potentials with the rather general asymptotic behavior:
\begin{equation}\label{V-IR}
V=\ell^{-2}V_\infty(\log\lambda)^{P} \lambda^{2Q},
\end{equation}
In the IR, the bulk potential should give rise to confining geometry characterized by the area law of the Wilson loop, which indicates a linear potential between two quarks. Requiring confinement, a mass gap and a discrete Regge-like glueball spectrum fixes the asymptotic form of the bulk potential in the IR to be (\ref{V-IR}) with  \cite{gkn}:  
\be 
Q=\frac 2 3, \qquad P=\frac 1 2.
\ee

The corresponding IR-regular solution of the superpotential is:
\begin{equation}
W(\lambda)\simeq W_\infty \ell^{-1}(\log\lambda)^{\frac P 2} \lambda^Q,\label{W-IR}
\end{equation}
where
\begin{equation}
W_\infty=3 V_\infty^{\frac 1 2}.
\end{equation}

\subsection{The non-perturbative scale} \label{npscale}

In four-dimensional Yang-Mills, the fundamental parameter that defines the theory is not the coupling constant, which depends on the energy scale, but rather the value of the non-perturbative RG-invariant scale $\Lambda$ which sets the scale  of all dimensionful observables. A choice of the value of $\Lambda$ is in one-to-one correspondence with an RG flow  trajectory, i.e. a particular solution of the  $\beta$-function equation,
\be\label{np1}
\mu {d\lambda  \over d\mu} = \beta(\lambda) \quad \Rightarrow \quad \mu = \mu_0 \exp \int^{\lambda(\mu)}_{\lambda(\mu_0)} {d\lambda \over \beta(\lambda)}\,.
\ee
A choice of $\Lambda$ is equivalent to a choice of initial condition $\lambda(\mu_0)=\lambda_0$ for equation (\ref{np1}).

In perturbation theory, we can define $\Lambda$ using e.g. the one-loop  $\beta$-function, for example, as the scale at which the coupling becomes infinite. Explicitly, using $\beta_{1-loop}(\lambda) = -b_0 \lambda^2$, and integrating equation (\ref{np1}), we obtain the usual  definition:
\be\label{np2}
\Lambda_{1-loop} = \mu \exp\left[-{1\over b_0\lambda(\mu)}\right] = \mu_0 \exp\left[-{1\over b_0\lambda_0}\right],
\ee
which is constant  along an RG-trajectory governed by  the one-loop beta-function, with initial conditions specified by $\lambda(\mu_0) = \lambda_0$. As we anticipated, the value of $\Lambda$ is in one-to-one correspondence with a choice of the coupling at a given scale $\mu_0$.
Of course, the definition (\ref{np2}) is not exact, and higher order terms in the $\beta$-functions will introduce a $\mu$-dependence in the right-hand side. To define a truly RG-invariant scale, we notice that the quantity
\be\label{np3}
\Lambda_{YM} = \mu \exp\left[-\int^{\lambda(\mu)}_{{\lambda}^\ast} {d \lambda \over \beta(\lambda)}\right]
\ee
is RG-invariant for any choice of a reference value ${\lambda}^\ast$. Changing this reference value only changes $\Lambda$ by a multiplicative constant, which we can fix by relating the expression (\ref{np3}) to the value of a physical observable, for example an RG-invariant operator. This will be done in the following sections\footnote{The  expression (\ref{np3})  is also  scheme-dependent, in that it depends on the choice of scheme in the $\beta$-function. But in any  scheme, we can relate $\Lambda$ to the (scheme-independent) value of a physical  observable of the theory.}. Again,  specifying a value of $\Lambda$ in (\ref{np3}) picks a special solution of the RG-equation (\ref{np1}), determined by the initial condition $\lambda(\Lambda)= {\lambda}^\ast$, i.e. it picks  a  unique physical theory.

On the gravity dual we have exactly the same situation: a  choice of the superpotential $W(\lambda)$ fixes the holographic $\beta$-function via equation (\ref{beta}), and an RG-trajectory is fixed by further specifying an initial condition for equation (\ref{sol1}). Given such a solution,  we can define a  quantity with the dimension of a mass scale,  which is  constant along the radial flow: first define the scalar function:
\be \label{A}
\A_{\phi^\ast}(\phi) = -{1\over 6}\int^\phi_{{\phi}^\ast} d\phi {W \over \de_\phi W}\,.
\ee
This quantity is a function of $\phi$ and of a reference point $\phi^\ast$, whose specific value affects $\A$ only by an additive constant.  

By equation (\ref{sol2}), when evaluated on a solution $\phi(u)$, $\A(\phi(u))$ coincides with the metric scale factor, up to an additive constant.   
 Therefore, the quantity:
\be\label{np4}
\Lambda = \xi \, {e^{A(u)} \over \ell} \exp[-\A_{{\lambda}^\ast}(\lambda(u))],
\ee
where  $\xi$ is an arbitrary dimensionless parameter, is $u$-independent  along any domain-wall solution like (\ref{sol2}). 
 
In terms of $\lambda =\exp[{\sqrt{3/8}}\phi]$, the function defined in  (\ref{A})   becomes:
\be \label{np4-1}
\A_{\lambda^\ast}(\lambda) = -{4\over 9}\int^\lambda_{{\lambda}^\ast} {d\lambda \over \lambda^2} {W \over \de_\lambda W} = \int^\lambda_{{\lambda}^\ast} {d\lambda' \over \beta(\lambda)},
\ee
where in the second equality we have used  the expression of the holographic $\beta$-function in terms of the superportential, equation (\ref{beta}). Recall also that $e^{A(u)}$ is identified in the holographic dictionary with the energy scale $\mu$. Thus, equation (\ref{np4}) reproduces, up to a normalization  constant, the field theory expression (\ref{np3}). Finally, if we take the perturbative approximation for $W(\lambda)$ in (\ref{W-UV}) and we stop at $O(\lambda^2)$, we recover the standard  one-loop expression (\ref{np2}).

 In order to completely  specify the holographic version of the field theory scale $\Lambda_{YM}$, we have to choose the constant $\xi$ and the reference point ${\phi}^\ast$ (these two parameters are in fact redundant,  since we can always reabsorb a shift in ${\phi}^\ast$ in a redefinition of $\xi$). For later convenience, we fix this ambiguity  by choosing:
\be \label{np5}
\xi = \left(\frac{4 M^3 \ell^3}{N_c^2}\right)^{\frac 1 4}, \qquad {\phi}^\ast = 0.
\ee
Since, as we discussed earlier, $M^3\sim N_c^2$, the above choice ensures that  $\Lambda$ stays finite in the $N_c\to \infty$ limit. As we will see, with this choice  $\Lambda$ matches  the  scale  of the glueball condensate.

With the definition (\ref{np5}), the non-perturbative scale  is specified by:
\begin{itemize}
\item Two  parameters entering in the bulk action, namely $M$ and $\ell \equiv \lim_{\phi\to -\infty}\sqrt{12/V(\phi)}$.
\item The parameter ${A}^\ast = A(\phi=0)$   specifying the initial condition for the bulk solution.
\end{itemize}
The product $M \ell$ can be fixed for example from the high-temperature regime of the theory \cite{thermo2}:
\be \label{np7}
(M \ell)^3 = {{N_c^2}\over 45 \pi^2}.
\ee

We finally arrive at the definition:
\be\label{np6}
\Lambda =  \xi \mu \,e^{-\A(\lambda(\mu))}= \xi {e^{{A}^\ast} \over \ell} , \qquad {A}^\ast \equiv A(\phi=0), \qquad \xi = \left({4\over 45\pi^2}\right)^{1/4}. 
\ee

This definition still  has a one-fold degeneracy in choosing the holographic model dual to a  Yang-Mills theory with a given scale $\Lambda$. However, notice that ${A}^\ast$ and $\ell$ are not separately observable, since $\ell$ ultimately affects only a choice of energy units. Therefore, for a given {\em physical} scale $\Lambda$, we are free to choose $\ell$ to be any reference scale, and fix  the integration constant ${A}^\ast$ to reproduce the physical value of $\Lambda$. For a thorough discussion about this point, see \cite{thermo3}.

\section{Nonperturbative gluon effective potentials}

\subsection{Holographic effective actions}

In a previous work \cite{kln}, a general formalism was presented
where the holographic RG flow and the quantum effective action are derived
from the gravitational dual.

In holography, the generating functional of the connected correlators is equal to  the action of the bulk theory evaluated on-shell. This is  divergent due to the infinite volume of the bulk,  and  these divergences  parallel the UV infinities of the dual  quantum field theory

As in field theory, one can define a finite, regularized generating functional in a cut-off version of the theory, by imposing  a (large-volume) cutoff on the  radial coordinate $u$. 

As  it was shown in \cite{papa,kln}, once a suitable IR regularity condition is imposed on the superpotential,   the regularized on-shell action is a UV boundary term which depends on   the scalar field and induced metric at the cutoff, and has the form:
\be\label{Sreg}
S^{(reg)}[\gamma_{\mu\nu}, \phi] = M^{3}\int_{u=u_{UV}} d^4x\,\sqrt{-\gamma}\left[W(\phi)- U(\phi)R
+\left({{W(\phi)}\over {W'(\phi)}} U'(\phi)\right) \frac 1 2 \gamma^{\mu\nu}\partial_\mu\phi\partial_\nu\phi\right],
\ee
where $u_{UV}$ is the position of the UV cutoff,  $W(\phi)$ is the superpotential, and the function $U(\phi)$ that enters the second-derivative terms is given by:
\be\label{Uphi}
U(\phi) = e^{-2\A(\phi)} \int_{\phi_{IR}}^\phi d{\psi}{1\over W'({\psi})} e^{2\A({\psi})},
\ee
with the  function $\A(\phi)$  defined as:
\begin{equation}
\A(\phi) = -{1\over 6} \int^\phi_{{\phi}^\ast} {{W(\psi) }\over{ W'(\psi)}} d\psi\label {A},
\end{equation}
where $\phi^\ast$ is a reference value of the scalar field and its choice amounts to fixing the integration constant in $\A(\phi)$, but it does not change equation (\ref{Uphi}). By equation (\ref{sol2}), when evaluated on the solution $\phi(u)$, the function $\A(\phi(u))$  conicides up to a constant with the scale factor $A(u)$ (in particular,  it has the same leading UV asymptotic behavior).

As discussed in
\cite{kln}, a way to translate the holographic cut-off into the dual field theory language is to notice that, in the homogeneous case,  the scale factor $e^{A(u)}$ of the induced metric corresponds to the field theory energy scale, and  the UV corresponds to the limit $e^A \to \infty$. If we cut-off the radial coordinate at a  point $u=u_{UV}$,  we can define the energy cut-off $\Lambda_{UV}$ by:
\be
\Lambda_{UV} = \ell^{-1} e^{A(u_{UV})} \equiv {1 \over \epsilon},
\ee
where in the last line we  have written the cutoff in terms of the small parameter $\epsilon$ typically used as cut-off for the conformal coordinate $r$  of Poincar\'e $AdS$, where $e^A \approx \ell/r$  in the UV.

In the UV, as $A\to \infty$, or $\lambda\equiv e^{\sqrt{3/8}\phi} \to 0$,  the functions $W$ and $U$ go to constant values,
\be
W = {6\over \ell}\left[1 + O(\lambda)\right] , \qquad U \simeq -{\ell \over 2}\left[1 + O(\lambda)\right], \qquad \lambda \approx 0.
\ee

On the other hand, writing the induced metric in the UV as
\be
\gamma_{\mu\nu} = e^{A(u)} \gamma^{(0)}_{\mu\nu}
\ee
in terms of  boundary metric $\gamma^{(0)}_{\mu\nu}$ which stays finite in the UV limit, the regularized on-shell action (\ref{Sreg}) becomes:
\begin{align}\label{Sreg2}
S^{(reg)}[\gamma_{\mu\nu}, \lambda] =& M^3 \ell^4 \int  d^4x\,\sqrt{-\gamma^{(0)}} \,\epsilon^{-4}\,W(\lambda)  \nonumber\\
& -M^3 \ell^2 \int  d^4x\,\sqrt{-\gamma^{(0)}}\,\epsilon^{-2}\,\left[ U(\lambda) R^{(0)}
-\left({W \de_\lambda U \over \de_\lambda W} \right) \frac 4 3 \gamma^{(0)\mu\nu} {\partial_\mu\lambda\partial_\nu\lambda \over \lambda^2}\right],
\end{align}
where the leading dependence of the cut-off is manifest, and we have written the  action as a function of the 't Hooft coupling $\lambda$ at the cut-off.

To obtain finite results once the UV cut-off is removed, one needs to perform the renormalization procedure.
As in a renormalizable theory, one identifies all the UV divergent terms
and subtracts them by adding a finite number of counterterms on the boundary.

In \cite{kln} an explicit expression was found for the renormalized generating functional depending on the induced d-dimensional metric and the scalar field,  up to two derivatives:
\begin{equation}
S^{(ren)}[\gamma_{\mu\nu},\phi]=M^{3}\ell^{-1}\int d^4x\,\sqrt{-\gamma}\left(Z_0(\phi)+ Z_1(\phi)R
+ Z_2(\phi) \frac 1 2 \gamma^{\mu\nu}\partial_\mu\phi\partial_\nu\phi\right),\label{running}
\end{equation}
where $\phi$ and $\gamma_{\mu\nu}$ are evaluated on an arbitrary slice in the bulk, and the coefficient functions are:
\begin{eqnarray}
Z_0(\phi)&=&D_0 e^{-4\A},\label{Z0}\\
Z_1(\phi)&=&D_0 G_0^{(1)}e^{-2\A}+D_1 \ell^2e^{-2\A}, \label{Z1}\\
Z_2(\phi)&=&\left(D_0 G_0^{(2)}+D_2\right)W'^{-2}e^{-2\A}+D_1 \ell^2 \frac W{W'}\left(e^{-2\A} \right)'. \label{Z2}
\end{eqnarray}

The constants $D_0,\,D_1,\,D_2$, which are dimensionless,  are determined by the difference between the UV-finite terms in (\ref{Sreg2}) and the corresponding terms in the counterterm action, which takes the same form as (\ref{Sreg2}) but with different functions $W(\phi)$ and $U(\phi)$  \cite{kln}.
The  constants $D_0,\,D_1,\,D_2$ completely parametrize  the scheme-dependence which arises from  choice of counterterms with up to two derivatives. We will discuss this point further in Section \ref{scheme}.
The function $\A(\phi)$ is defined in equation (\ref{A}). A change in the reference point ${\phi}^\ast$ in (\ref{A})   can be reabsorbed into a redefinition of $D_0,D_1,D_2$.

Finally, the functions $G_0^{(1)}(\phi)$ and $G_0^{(2)}(\phi)$ appearing in (\ref{Z1}-\ref{Z2})  are defined as:
\begin{align}
G_0^{(1)}( \phi)=& G_0^{(1)}( \phi^\ast)+\frac 1 {3}\int_{\phi^\ast}^\phi d\tilde \phi \,e^{-2\A}\,W'^{-2}\Big(2WU'-W'U\Big), \label{G1} \\
G_0^{(2)}( \phi)=& G_0^{(2)}(\phi^\ast)+2\int_{\phi^\ast}^\phi d\tilde \phi \,e^{2\A}\,W'\left[\left(e^{-4\A}\right)'' \frac W {W'}U'+\left(e^{-4\A}\right)' \left( \frac W {2W'}U'\right)'\right] \nonumber\\
&\!\!\!\!\!\!\!-\frac{5}{3}\int_{\phi^\ast}^\phi d\tilde \phi \,e^{-2\A}\,W U'+2\int_{\phi^\ast}^\phi d\tilde \phi \,e^{2\A}\,W'^2\left(G_0^{(1)}e^{-2\A}\right)' ,\label{G2}
\end{align}
where the integration constants $G_0^{(1)}( \phi^\ast)$ and $G_0^{(2)}( \phi^\ast)$ are chosen in such a way that:
\begin{equation}
G_0^{(1)}( \phi_{UV})=G_0^{(2)}( \phi_{UV})=0.
\end{equation}

From (\ref{Sreg}) and (\ref{running}), one  can derive  bare vacuum expecation values in the cut-off theory, or the renormalized vacuum expectation values, respectively. In general, we will deal with operators whose source is   a function  $J(\phi)$, rather than $\phi$ itself. The corresponding (bare and renormalized) expectaction values are:
\be \label{vev}
\<\O\>^{(bare)}(\phi,\epsilon) = \frac{1}{\sqrt{-\gamma^{(0)}}} \frac{\delta S_{\epsilon}[\phi,\gamma^{(0)}_{\mu\nu}]}{\delta J(\phi)}, \qquad \<\O\>^{(ren)}(\phi,\mu) = \frac{1}{\sqrt{-\gamma^{(0)}}} \frac{\delta S^{(ren)}[\phi, \gamma^{(0)}_{\mu\nu}]}{\delta J(\phi)}.
\ee
The first will depend on the coupling at the cutoff, and also explicitly on the cutoff scale; the second is finite as the cut-off is removed, and will depend on the choice of the coupling at the fixed renormalization scale $\mu$
set by the scale factor in $\gamma_{\mu\nu}$.

The 1PI  action, or quantum effective action $\Gamma(\<\cO\>)$ is obtained  by Legendre transforming the renormalized generating functional (\ref{running}) with respect to the  source  $J$. We can compute the effective action in the regularized theory starting from (\ref{Sreg}), or  the renormalized effective action starting from (\ref{running}). In either case, we use the definition\footnote{We will denote the argument of the effective action $\Gamma$ by $\O$ omitting the brakets. We will reserve the notation $\<\O\>$ to indicate the physical vev in vacuum (i.e. in the absence of sources).}: 
\be
\Gamma[\cO,\gamma] =\int d^d x \sqrt{- \gamma^{(0)}} J(\O)\, \O -S(\O),
\ee
where $J$ is written as a function  of $\O$ by inverting equation (\ref{vev}). The physical vacuum expectation value for vanishing source is found, as usual, by extremizing the effective action:
\be
{\de \Gamma \over \de \O}\Big|_{\O = \<\O\>} = 0.
\ee

In the following sections, we will separately discuss the bare and renormalized effective actions for two different operators, i.e. the Yang-Mills  operator $Tr F^2$ and its renormalization-group invariant version,
$\big({-\beta(\lambda) }/{2\lambda^2}\big)Tr F^2$, which coincides with the trace of the stress tensor
via the conformal anomaly. Before we enter the detailed discussion, we make a few remarks about the general structure of  effective actions obtained from the above procedure.

 To lowest order in derivatives the vev of the {\em renormalized} dual operator is obtained by the variation of the zero-derivative term in the generating functional (\ref{running}) with respect to the  source $J$:
\begin{equation}
\<\cO(x)\>_J=\frac 1{\sqrt{- \gamma^{(0)}}} \frac {\delta S^{(ren)}}{\delta J(x)}=M^3\ell^{-1}\frac {d Z_0(\phi)}{d J}\label{vev-Z0}
\end{equation}
where $Z_0$ is in \eqref{Z0}.

This receives derivative corrections from the derivative  terms in (\ref{running}). However, as shown in \cite{kln},
the general expression for the  renormalized effective action up to two derivative order is rather simple:
\begin{align}
\Gamma[\cO,\gamma] =&\int d^d x \sqrt{- \gamma^{(0)}} J {\mathcal O} -S^{(ren)} \nonumber\\
=&\int d^d x \sqrt{- \gamma^{(0)}}\,\left[{\mathcal O} J_0({\mathcal O})-M^3\ell^{-1}Z_0(J_0({\mathcal O}))\right]\nonumber\\
&-M^3\ell^{-1}\int d^d x\sqrt{- \gamma^{(0)}}\Big[Z_1(J_0({\mathcal O}))R+\phi_0'^2({\mathcal O}) Z_2(J_0({\mathcal O}))\frac 1 2\gamma^{\mu\nu}\partial_\mu {\mathcal O}\partial_\nu  {\mathcal O} \Big]\label{1PI}
\end{align}
where $J_0(\mathcal O)$ is the inverse function of (\ref{vev}) {\em determined at zeroth-order only}, i.e. it  coincides with the zero-derivative term of the full source $J(\mathcal O)$ as a function of the vev $\<\cO\>$.

In other words, the derivative corrections to (\ref{vev}) are cancelled by similar two-derivative terms in $S^{(ren)}$, and (\ref{1PI}) is the complete result to two-derivative order. In  Section \ref{effective}, we apply this to the compution of the effective action of gluonic operators ${\cal G}$ and ${\cal T}$ given respectively by  (\ref{G}) and (\ref{T}).

\subsection{The subtraction scheme} \label{scheme}

Before extracting explicit results for the glueball operator effective potential in QCD-like models, in this section we will give a few details about the subtraction procedure we use to obtain the renormalized effective action, and how it may relate to  schemes adopted in standard quantum field theory.
General renormalisation of Einstein-Dilaton theories was developed in \cite{papa}.
Here, we will try to answer as explicitly as possible the question,
{\em what is the subtracted part of the generating functional in the holographic scheme, in terms of physical quantities that one can relate to the field theory?}.

We will concentrate on the zero-derivative term in the action, whose subtraction corresponds to an additive renormalization of the vacuum energy.
Similar considerations apply to the two-derivative terms, which correspond to
a renormalization of the Einstein-Hilbert term and of the kinetic term associated to a space-dependent coupling.

At the zero-derivative order, the regularized generating functional  is the vacuum energy in flat spacetime, calculated in the cut-off theory  and it is given by the first term in equation (\ref{Sreg2}):
\be \label{Sreg0}
S^{(reg)}_0 = M^3 \ell^4 \int d^4 x {W(\lambda_\epsilon) \over \epsilon^4}.
\ee
This expression is constructed by first choosing a solution $(A(u), \lambda(u))$ of the homogeneous equations of motion (\ref{sol1}) , with the superpotential and all the integration constants fixed. Then,  $\epsilon$ and $\lambda_\epsilon$ are defined
by the scale factor and dilaton on an UV slice $u_{UV}$:
\be \label{lambda-cutoff}
\epsilon = \ell \, e^{-A(u_{UV})}, \qquad \lambda_\epsilon = \lambda(u_{UV}).
\ee
Finally, one evaluates the superpotential $W(\lambda)$  of the solution at $\lambda_\epsilon$ in equation (\ref{Sreg0}).

The energy cut-off $\Lambda_{UV}$  in the holographic scheme is related to $\epsilon$ by
\be \label{LUV}
\Lambda_{UV} \equiv 1/\epsilon = \ell^{-1} e^{A(u_{UV})}.
\ee

Notice that, in the definitions above,  we could have avoided any reference to the UV value of the conformal coordinate $u_{UV}$: the coordinate-invariant data are, in a given solution, the value $\lambda_\epsilon$ of the dilaton {\em when the scale factor takes on a given value $1/\epsilon$}. Similarly, since in the coordinates (\ref{metric}), the warping factor $A$ enjoys a shift symmetry, the only invariant way to decide whether the cut-off is actually in the UV is whether {\em the value of the coupling at the cut-off is small} since, in the UV-complete solution, $\lambda \to 0$ as $e^A \to \infty$. Thus, at the cut-off we choose
\be
\lambda_\epsilon \ll 1 .
\ee
This is similar to the situation in  ordinary QCD, where the theory is taken to be perturbative, and the coupling small, at the UV cut-off scale. As we will see below, as in QCD, this implies a parametric hierarchy between the UV cut-off  and the non-perturbative scale  $\Lambda$ defined in Section \ref{npscale}, $\Lambda/\Lambda_{UV} \ll 1$.

One can change the cutoff by changing $\epsilon$ and $\lambda_\epsilon$ at the same time, by following the flow of the chosen solution $(A(u),\lambda(u))$. The limit $\epsilon \to 0$ corresponds to removing the cut-off, and can be taken only after a subtraction is performed: in fact,   $W(\lambda)$  has a finite value at $\lambda=0$, thus  by equation (\ref{W-UV}) the regularized vacuum energy (\ref{Sreg0}) diverges as
\be
S_0^{(reg)} \approx \int d^4x \,  6 (M \ell)^3 \Lambda_{UV}^4 .
\ee

To obtain the renormalized on-shell action one has to add boundary counterterms at the UV slice. At the zero-derivative order, the appropriate covariant counterterm is \cite{bianchi,papa,kln}:
\be\label{Sct}
S_0^{ct} = -M^3  \int_{u_{UV}} d^4 x \, \sqrt{\gamma}\,  W^{ct}(\lambda) = -M^3 \ell^4  \int d^4 x \, {W^{ct}(\lambda_\epsilon) \over \epsilon^4},
\ee
where $W^{ct}(\lambda)$ is an arbitrary solution of the superpotetial equation (\ref{superpotential-eq}) which flows to the same UV-AdS fixed point, and in the last equality we have used $\sqrt{\gamma} = e^{4A(u_{UV})} = (\ell /\epsilon)^4$.

Choosing $W^{ct} \neq W$ will subtract the divergence but leave a non-zero finite result. In fact,  as discussed for example in \cite{thermo2,KN,kln}, for small $\lambda$ {\em any} solution of equation (\ref{superpotential-eq}) takes the form:
\be\label{wtot}
W(\lambda) = W_0 (\lambda) + C W_1 (\lambda) + \ldots
\ee
where   $C$ is an arbitrary real number  and $W_0(\lambda)$ and $W_1(\lambda)$ are universal functions, and the order of the subleading terms will be  specified later. In particular, if the bulk potential $V(\lambda)$ has an analytic expansion around $\lambda=0$, then  $W_0$ has an analytic expansion in integer powers of $\lambda$,
\be\label{w0}
W_0(\lambda) = \sum_{n=0}^{+\infty} w_n \lambda^n,
\ee
with all coefficients $w_n$ determined by the expansion coefficients of $V(\lambda)$ around $\lambda=0$ \cite{thermo2} (the first two terms are given explicitly in equation (\ref{W-UV}).

The function  $W_1$ is determined by $W_0$ and is  given by:
\be\label{w1}
W_1(\lambda) = \exp\left[-{16\over 9}\int^\lambda {d\lambda' \over\lambda^{'2}} {W_0\over \de_{\lambda'} W_0}\right].
\ee
The above equation defines  $W_1$ up to an integration constant, but given that $C$ in \eqref{wtot} is arbitrary we can choose this constant at will.

Using the power-law expansion for $W_0(\lambda)$ in equation (\ref{w1}) it is easy to  obtain explicitly the small-$\lambda$ expansion for $W_1$: this consists in a power series similar to (\ref{w0}),  multiplied by an overall non-analytic factor:
\be\label{w1-3}
W_1(\lambda) = e^{-{4 \over b_0 \lambda}} \lambda^{4b_1 \over b_0^2}\sum_{n=0 }^{+\infty}\tilde{w}_n \lambda^n,
\ee
where
\be
b_0 \equiv {9\over 4} {w_1\over w_0}, \qquad b_1 \equiv  {9\over 4} {w_1\over w_0}\left( {w_1\over w_0} -  2{w_2\over w_1}\right)
\ee
are the first two beta-function coefficients \cite{thermo2}, and the power series  coefficients $\tilde{w}_i$ are determined by $w_i$ except for a common overall factor. One can go further in (\ref{wtot}) by adding the subleading order terms: these are proportional to $C^2$ and to the square of the non-analytic exponential in $\lambda$ \cite{KN}.

It is instructive to  compare  equation (\ref{w1}) with equations (\ref{np4}-\ref{np4-1}): we see that, up to the choice of a multiplicative constant which we can choose   by rescaling $C$ to be related to  $\xi$ defined in (\ref{np5}), we have:
\be\label{w1-2}
W_1 (\lambda(u_{UV})) = {e^{-4A(u_{UV})}\over \xi^4} (\ell \Lambda)^4 = {(\epsilon  \Lambda)^4 \over \xi^4} , \qquad \xi \equiv \left(\frac{4 M^3 \ell^3}{N_c^2}\right)^{\frac 1 4}.
\ee

Thus, at the cut-off surface, we can write both $W(\lambda_\epsilon)$ in (\ref{Sreg0})  and $W^{ct}(\lambda_\epsilon)$   in (\ref{Sct}) in the same form (\ref{wtot}),  the only difference  being encoded in the contants $C,C^{ct}$ multiplying $W_1$:
\be\label{ws}
W = W_0(\lambda_\epsilon) + C \,\left({\epsilon \Lambda \over \xi}\right)^4 + O(\epsilon^8), \qquad W^{ct} = W_0(\lambda_\epsilon) + C^{ct} \,\left({\epsilon \Lambda \over \xi}\right)^4 + O(\epsilon^8),
\ee
where the $O(\epsilon^8)$ terms are corrections to the leading terms in (\ref{wtot}).

From (\ref{Sreg0}) and (\ref{Sct}) we see that the renormalized action is:
\be
S^{(ren)} =N_c^2 \int d^4x \, \ell\, (C-C^{ct}) {\Lambda^4 \over 4} + O(\epsilon^4),  
\ee
The coefficient $D_0$ appearing in \eqref{Z0} is then given by 
\be
D_0=C-C^{ct}.
\ee
Thus, holographic renormalization of the vacuum energy proceeds via  an {\em additive} renormalization  by a function of the cut-off (including divergent and finite terms):
\be
S^{ct}(\epsilon) = M^3 \ell^4 \int  {W(\lambda_\epsilon) \over \epsilon^4} \, + \,  {\Lambda^4 \over 4}  + O(\epsilon^4)
\ee

In the above expression,  the dependence on the cut-off is both in the explicit $1/\epsilon^4$ and in the dependence through $\lambda_\epsilon$.
We can  take one step further and express the subtracted function of the cutoff purely in terms of physical quantities. Although a close analytic expression cannot be obtained,  this can be done explicitly order by order in a log-expansion, as we show below.

In the UV, we can express $\lambda$ as a function of the cut-off $\epsilon$  and the nonperturbative scale $\Lambda$, by using   equations  (\ref{np4}) and  (\ref{np4-1}): for small $\lambda$,
\be
\A(\lambda) = {1\over b_0 \lambda} \;+\; subleading.
\ee
Evaluating this expression at the cut-off $u_{UV}$, using equation (\ref{np4}), recalling that $A(u_{UV}) = \log (\ell/\epsilon)$ and neglecting subleading terms\footnote{The most important ones are a logarithmically divergent term and a constant term in the limit $\lambda\to 0$. Further terms vanish as  powers of $\lambda$.}, we have the approximate relation:
\be\label{lambdaeps}
\lambda_\epsilon = {1\over b_0 \log (\xi \Lambda_{UV}/\Lambda )} \left[1 + O \left({\log [\log  \Lambda_{UV}/\Lambda ]\over \log\Lambda_{UV}/\Lambda}\right)\right],
\ee
where $\xi$ is defined in equation (\ref{np5}) and we have replaced $\epsilon$ by $\Lambda_{UV}$ using  (\ref{LUV}).

Thus, using equation (\ref{lambdaeps}) we can have an  expression, written as a power series,
for the function of the cut-off that we are using in the subtraction, in terms of the phyisical parameters, i.e. the cut-off scale $\Lambda_{UV}$ and the RG-invariant non-perturbative scale $\Lambda$:

\be\label{Sct2}
S^{ct}(\Lambda,\Lambda_{UV}) = (M \ell)^3 \Lambda_{UV}^4 \sum_n  (\ell w_n) \left[\log (\xi \Lambda_{UV}/\Lambda)\right]^{-n}\Big(1 + \ldots\Big)  + N_c^2(\ell C^{ct}){\Lambda^4  \over 4} + O\left(\Lambda^8 \over \Lambda_{UV}^4\right)
\ee
In this expression, the divergent terms are uniquely determined by the expansion coefficients $w_n$ of the leading superpotential (\ref{w0}), which are in turn uniquely determined by the expansion coefficients of the bulk potential $V(\lambda)$ around $\lambda=0$.  The same can be said for the
universal subleading terms, which correct each term in the series at $O(\log\log (\Lambda_{UV}/\Lambda) \,/\, \log  (\Lambda_{UV}/\Lambda))$. The finite, non-universal term depends only on an overall coefficient $C^{ct}$ and determines the renormalized vacuum energy in terms of the non-perturbative scale.

Equations (\ref{ws}) and (\ref{Sct2})  make the subtraction scheme manifest in terms of 1) the physical quantities, i.e. the cutoff scale $\Lambda_{UV}$ and the non-perturbative scale $\Lambda$ associated to the solution; 2) a scheme-dependent constant $C^{ct}$, which determines completely the choice of the counterterm.

As a final remark, notice that, by equation (\ref{lambdaeps}),  requiring $\lambda$ to be small at the cutoff is the same as requiring that
\be
{\Lambda \over \Lambda_{UV}} \ll 1,
\ee
which is the usual condition on the separation of the UV scale from the IR  scale for QCD to be perturbative at the cutoff.

\subsection{Effective actions for glueball operators} \label{effective}

There are several different composite operators that one can associate to the Yang-Mills field strength. They have different effective actions, which arise by Legendre-transforming the generating functional with respect to
different source functions $J(\phi)$.  We will consider the following choices.

\begin{enumerate}
\item One of them is simply ${\cal G} \equiv Tr F^2$. As we discuss below, this is  obtained by choosing  the source
$J = {-1/{2\lambda}}$.
Since the generating functional is RG-invariant, but the coupling is scale-dependent, this  operator is itself non-RG-invariant. Moreover,
in the holographic theory it depends on the relation between the field theory 't Hooft coupling and the bulk field $\lambda$, which may be non-trivial in the IR.

\item
The RG-invariant version of the  gluon  composite operator is
\be
{\cal T}=- {\beta(\lambda)\over 2\lambda^2} Tr F^2 .
\ee
This coincides with the trace of the stress-tensor, and as we will see in Section \ref{invariant} its vev  can be obtained by
taking the source to be proportional to the function $\A(\phi)$ in (\ref{A}), i.e. the scale factor. This operator is universal, and its potential does not depend on the identification of the t'Hooft coupling in the bulk.

\item The VEVs and effective actions of  the operators above can be obtained in both the bare regularized theory and the renormalized theory, depending whether one takes (\ref{Sreg2}) or (\ref{running}) as a starting point.

\item Including second order derivative terms in the effective action, we may also consider redefinitions of the above operators such that they are   canonically normalized.
In the case of  renormalized operators, by equation (\ref{1PI}) we see that the kinetic term has the universal form:
\be
{\phi_0'}^2(\O) Z_2(\phi_0 (\O)) (\de_\mu O)^2 =  Z_2(\phi_0(\O)) (\de_\mu \phi_0(\O))^2,
\ee
where $\phi_0(\O)$ is the zeroth-order relation between $\phi$ and $\O$ obtained from (\ref{vev}) and $Z_2(\phi)$ is given in equation (\ref{Z2}). Thus, the canonically normalized operator $\tilde{\O}$ is defined by the relation
\be\label{vev-ii}
\sqrt{Z_2(\phi)}\, d\phi = d\, \O^{(c)}.
\ee
Therefore, canonical normalization is defined through a universal function $\O^{(c)}$, independently of the initial choice for the operator $\O$.  However, the potential for  $\O^{(c)}$ does depends on the original definition of the operator, through the  source function $J_0(\O)$ in (\ref{1PI}).

The same consideration applies to the bare, regularized operator, in which case one has to substitute $Z_2(\phi)$ with the function $W U'/ W'$, see \eqref{Sreg2}.
\end{enumerate}

\subsubsection{The renormallized  composite gluon operator $Tr F^2$} \label{secTrF2}

The Lagrangian of Yang-Mills theory is
\begin{equation}
\mathcal L_{YM}= -\frac 1 {2g_{YM}^2} Tr[ F^2]=- \frac{N_c}{2\lambda}Tr[ F^2],
\end{equation}
where $ \lambda =  g_{YM}^2 N_c$ is the 't Hooft coupling. In the following, we will absorb the factor $N_c$ into the normalization of the vector potential and define the operator coupled to the 't Hooft coupling as the YM operator
\begin{equation}
\mathcal L_{YM}=- \frac{1}{2\lambda} {\cal G}.
\end{equation}
Thus, expectation values of (products of) ${\cal G}$ can be obtained by taking functional derivatives of the quantum generating functional with respect to $J=(-4\lambda)^{-1}$.

We can then make the coupling space-time dependent and use it as a source to perform the Legendre transform: 
\begin{equation}
\tilde\Gamma[{\cal G}]=\int d^4 x \sqrt{- \gamma^{(0)}} J(x){\cal G}(x) -S^{(ren)},\qquad {\cal G}(x)=\frac 1 {\sqrt{- \gamma^{(0)}}}\frac {\delta S^{(ren)}}{\delta J(x)},
\end{equation}
where $J(x)=[-2\lambda(x)]^{-1}$. The minimum of this effective potential is in the far infrared where both the vev $\<{\cal G}\>= \<Tr F^2\>$ and 't Hooft coupling $\lambda$ go to infinity:
\be\label{Ginf}
\<{\cal G}\>_{min}\rightarrow \infty, \qquad \lambda_{min} \rightarrow \infty,
 \ee
 which sets the source $J_{min}\sim \lambda^{-1}_{min}$ to zero. Therefore, after a deformation around the UV fixed point, the vev will flow to the minimum of the effective potential in the strongly coupled IR. In this standard effective potential, one can only deduce the properties of the IR physics, not the information in the intermediate scale between the UV and IR.

To study the RG scale dependence of the effective potential,  it is more appropriate to use  a modified version of the Legendre transformation in which the source, instead of being the full coupling, is taken to be the fluctuations of the coupling  around a background value. 

In a QCD-like  model, at any finite intermediate scale $\mu$, the background 't Hooft coupling $\bar\lambda(\mu)$ is finite, thus the background source $\bar J$ is non-vanishing. In this modified computation, we will perform the Legendre transformation with respect to the fluctuation $\tilde J(x)=J(x)-\bar J(\mu)$, namely the difference between the full coupling $J(x)$ and the background coupling $\bar J(\mu)$ at the RG scale $\mu$. The effective action is given by 
\begin{equation}\label{Legendre}
\Gamma[{\cal G},\mu]=\int d^4 x \sqrt{- \gamma^{(0)}} \tilde J(x) \frac {\delta S^{(ren)}}{\delta \tilde J(x)} -S^{(ren)},
\end{equation}
where $\tilde J(x)=J(x)-\bar J(\mu)$ and $\bar J=(-2\bar\lambda)^{-1}$. Using this  definition, the minimum of the effective potential (the zeroth order term in the effective action $ \Gamma[{\cal G},\mu]$ ) will correspond to  a finite vev at the RG scale $\mu$ corresponding to  a finite coupling $\bar{\lambda}(\mu)$:
\begin{equation}
\<{\cal G}\>_{\min}=\<{\cal G}\>_{\lambda(x)=\bar \lambda} = \left.{\delta S^{(ren)}[J(x)] \over \delta J(x)}\right|_{\lambda(x) = \bar{\lambda}(\mu)}. 
\end{equation}
Due to the scale dependence of the background value, both the quantum effective action and the expectation values will depend on the renormaliztion scale $\mu$.
 
Now let us compute the full effective action from the Legendre transformation \eqref{Legendre}. To zero-derivative order, the YM operator ${\cal G}=Tr F^2$, as a function of $\mu$ and $\lambda$, is computed by the functional derivative of the renormalized generating functional \refeq{running} with respect to the 't Hooft coupling
 \begin{equation}
 {\cal G}[\mu,\lambda]= \frac {\delta S^{(ren)}}{\delta (- 2\lambda)^{-1}}=-2D_0\, N_c^2\xi^4 \big(\mu e^{- \A(\lambda)}\big)^4 \frac {\lambda^2}{\beta(\lambda)}+O(\partial^2), \qquad \xi \equiv \left(\frac{4 M^3 \ell^3}{N_c^2}\right)^{\frac 1 4}.
\end{equation}
In this calculation, we have identified the energy scale $\mu$ with the scale factor of the induced metric as in \eqref {mu}. The vev calculated here is the same as the standard definition because the variation of $\tilde J(x)$ coincides with that of the full coupling $J(x)$.

To derive the full coupling $\lambda({\cal G})$ as a function of the vev, one first inverts the relation between $ \lambda$ and $\<{\cal G}\>$
\begin{equation}\label{lambda-g}
\lambda=\lambda\left[\frac {\cal G}{\mu^4}\right].
\end{equation}
This will be used in the computation of the Legendre transformation.

In the Legendre transformation, we also need to know the scale dependence of the background 't Hooft coupling $\bar\lambda(\mu)$. Besides the RG scale, there is an additional dimensional parameter, the non-perturbative scale \refeq{np4}
\be
\Lambda= \xi\,\mu \,e^{-\A(\bar\lambda)},
\ee
which determines the specific RG flow under consideration.

Inverting the relation between $\bar\lambda$ and $ \mu /\Lambda$, one obtains the background 't Hooft coupling as a function of the energy ratio $\Lambda / \mu$ and thus the RG scale dependence of the background coupling
\be\label{lambda-mu}
\bar\lambda =\bar\lambda \left[\frac \Lambda\mu\right].
\ee
Using the inverse functions \eqref{lambda-g} and \eqref{lambda-mu} computed in the above ways and the definition  of the effective action \eqref{Legendre}, one can derive the effective action as a functional of the vev $\<{\cal G}\>$, the non-perturbative scale $\Lambda$ and the RG scale $ \mu$:
\be
\Gamma\left[{\cal G}, \mu, \Lambda \right]=-\int d^4x \, {1\over 2\lambda\left({\cal G}/\mu^4\right)} {\cal G} -  S^{(ren)} \left[\lambda\left({\cal G}/\mu^4\right)\right] + \int d^4x \, {1\over 2\bar{\lambda}\left(\Lambda/\mu\right)}{\cal G} 
\ee

We will present here the analytic results for the UV and IR, where one can expand the superpotential for small and large $\lambda$ respectively. We mean by the UV limit that the RG scale $\mu$ is much larger than the other dimension-one quantities
\be
\text{UV:}\qquad\mu\gg {\cal G},\qquad \mu \gg \Lambda,
\ee
while the IR limit indicates a hierarchy between the RG scale $\mu$ and the other dimension-one quantities in the opposite way
\be
\text{IR:}\qquad\mu \ll {\cal G},\qquad \mu \ll \Lambda.
\ee
The details are presented in Appendix A.

Using the asymptotic form of the superpotential in the UV and IR, we calculate the effective action in these two limits. The UV effective action is given by
\be
\Gamma_{UV}[{\cal G},\Lambda] =\int d^4 x\,  \left[ \frac {b_0}{8} {\cal G} \left( \ln\frac{{b_0 \cal G}}{2N_c^2 \Lambda^4}  -1 \right)
+\frac {b_0^{\frac 5 2}\xi^2 N_c}{128\sqrt 2}\, {\cal G}^{-\frac 3 2} (\partial{\cal G})^2\right], \quad \label{trFF-UV}
\ee
where we used the parameter $\xi$  defined in equation (\ref{np5}), i.e.
\be
\xi \equiv \left(\frac{4 M^3 \ell^3}{N_c^2}\right)^{\frac 1 4},
\ee
In order to fix all the coeficients we have chosen a renormalization scheme where the constants $D_i$ in the renormalized generating functional \eqref{running}, (\ref{Z0}-\ref{Z2}) take the following values:
\be\label{Dis}
D_0=1,\quad D_1=0,\quad D_2=\left(\frac{11} {24\pi^2}\right)^{2}.
\ee
This choice of $D_0$, as we will  see  in the next subsection, sets the minimum of the effective potential of the RG-invariant glueball operator at the value $\Lambda$. The definition of canonically normalized operator depends on the coefficient of the kinetic term and thus $D_2$. The value of  $D_2$ was chosen for convenience to simplify the coefficients in the effective action. In the numerical calculation of section 4, we will use another value of $D_2$ to set the minimum of the effective potential of the canonically normalized operator at the value $\Lambda$. A non-zero value for $D_1$ would make the Ricci scalar enter the action  second order in derivatives, which would then mix with the scalar kinetic term.  Since we are mostly interested in a Minkowski background, we have chosen to set its coefficient to zero.  

Now we can canonically normalize the operator  according to the kinetic terms:
\begin{equation}
{\cal G}^{(c)}=\frac {{b_0}^{\frac {5} {4}}N_c^{\frac 1 2}\xi} {2\sqrt 2}\, {\cal G}^{\frac {1}{4}}.
\end{equation}

The effective action of the canonically normalized vev ${\cal G}^{(c)}$ in the UV is
\be\label{EP-G-CN}
\Gamma_{UV}[{\cal G}^{(c)}]=\int \left\{ \frac {4{{\cal G}_{(c)}}^4}{b_0^{4}\,N_c^2\xi^4}\Big[ \ln \Big(\frac {{16\,{\cal G}_{(c)}}^4}{b_0^4\,N_c^4\xi^4\Lambda^4}\Big) -1\Big]
+ \frac 1 2\eta^{\mu\nu}\partial_\mu {{\cal G}^{(c)}}\partial_\nu {{\cal G}^{(c)}} \right\}, 
\ee
where we have neglected subleading terms in the potential.

In the IR $(\mu\ll {\cal G},\, \mu\ll\Lambda)$, the effective action reads
\bea
\Gamma_{IR}[{\cal G},\Lambda,\mu]=\int d^4x  &&\Bigg\{ \frac 1 2\xi^{\frac 3 {2}}\left(\frac 2 3\right)^{\frac {3}{8}}\left(\frac \Lambda \mu\right)^{-\frac 3 2}\Big(\ln\frac \Lambda \mu\Big)^{-\frac 3{8}}\,{\cal G} \nonumber\\
&& + \frac 1 2 \, E_1\,\mu^2\,\left(\partial\left[\left(\frac {\cal G}{\mu^4}\right)^{\frac 2{11}}\left(\ln\frac {\cal G}{\mu^4}	\right)^{-\frac {3}{44}}	\right]\right)^2 \Bigg\},	
\eea
where: 
\be
E_1= \left[\Big(4\Big)^{\frac {15}{11}}\left(\frac 2 3\right)^{-\frac {26}{11}}
\left(\frac {11}3\right)^{\frac 3{22}}\right]  \,\frac{\ell^2}{3}\,\xi^{\frac {28}{11}} G_0^{(1)}(\phi_{IR}), 
\ee
and $G_0^{(1)}(\phi_{IR})$ is the IR ($\lambda \to \infty$) limit of the function $G_0^{(1)}(\phi)$ defined in equation  (\ref{G1}).  
In the above expressions  we have fixed the parameters in the IR bulk potential (\ref{V-IR})  to the values appropriate for Yang-Mills theory, i.e.
\be
Q=\frac 2 3,\qquad P=\frac 1 2, 
\ee
Results for generic values of $Q$ and $P$ can be found in the Appendix. 

The canonically normalized operator ${\cal G}^{(c)}$ is determined by the kinetic term
\be
{\cal G}^{(c)}= E_1^{\frac 1 2}\, \mu\Big(\frac {\cal G}{\mu^4}\Big)^{\frac 2{11}}\Big(\ln\frac {\cal G}{\mu^4}	\Big)^{-\frac 3{44}}.
\ee

The corresponding effective action for the canonically normalized operator in the IR reads
\begin{align}
\Gamma[{\cal G},\mu,\Lambda]=\int\left[ E_2\left(\ln\frac \Lambda \mu\right)^{-\frac 3{8}} {{\cal G}_{(c)}}^{4}\left(\frac {{\cal G}^{(c)}}\Lambda\right)^{\frac 3 2}\left(\ln \frac {{\cal G}^{(c)}}{\mu}\right)^{\frac 3{8}}
+ \frac 1 2\eta^{\mu\nu}\partial_\mu{\cal G}^{(c)}\partial_\nu{\cal G}^{(c)}\right],
\end{align}
where we have defined: 
\be
E_2=\frac 2 3\left({16 \over N_c^{2}\xi^{4}}\right)^{\frac {11} {8}}\left[12\ell^{2} G_0^{(1)}(\phi_{IR})\right]^{-\frac {11} {4}}. 
\ee

After presenting the analytic results in the UV and IR where perturbative expansions are possible, we are going to compute the  full non-perturbative potential by numerically solving the equations. As an illustration, we consider a bulk scalar potential
\be\label{sim-pot}
V(\lambda)=\frac {12}{\ell^2}\left\{1+\frac {11}{27\pi^2}\lambda+\frac 1 {100}\lambda^{\frac 4 3}\Big[\log(1+\lambda)\Big ]^{\frac 1 2}\right\},
\ee
which has the correct UV and IR asymptotic behaviour given by \eqref{V-UV} and \eqref{V-IR} with $V_\infty=\frac 3 {25}$. The $1/100$ factor in front of the third term is to disentangle the UV behaviour from that of the IR.

\begin{figure}[h!]
\begin{center}
\includegraphics[width=14cm]{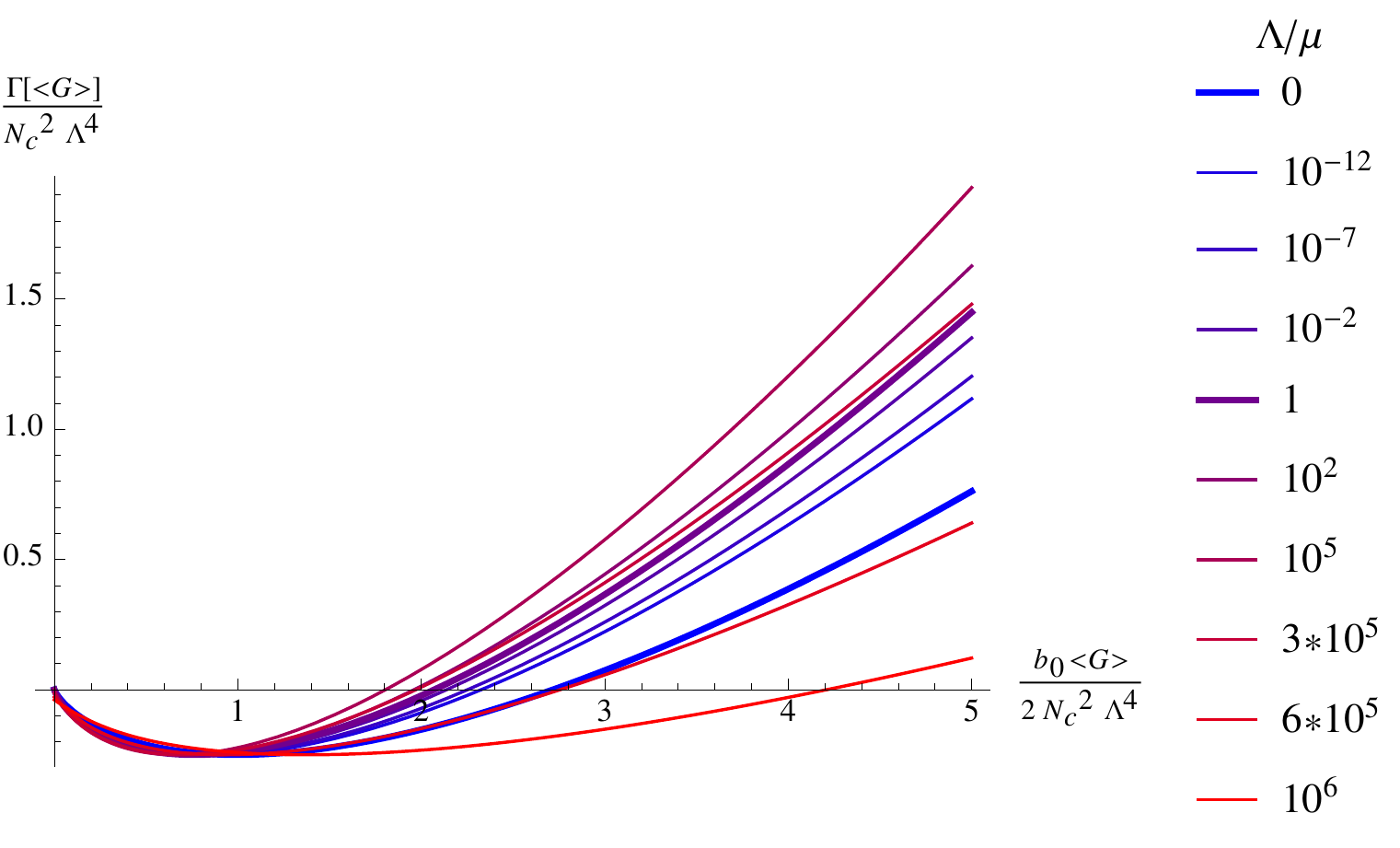}
\caption{The non-perturbative effective potential of $\cal G$. In the UV limit $\mu\rightarrow \infty$, the potential approaches the analytic form (3.56)
derived from the UV expansion. For large $\cal G$ in the IR limit $\mu\rightarrow 0$, the potential has a linear dependence on the vev $\cal G$ in accord with the IR expansion result. As we decrease the RG scale $\mu$, the potential goes up in the UV region, then slows down the trend of going up, and finally goes down in the IR region.}
\label{simp-pot-UV-IR}
\end{center}
\end{figure}

We present the result of the numerical calculation of the effective potential of $\cal G$ in figure \ref{simp-pot-UV-IR}.
In the extreme  UV region, the potential approaches the analytic form \eqref{trFF-UV} in the limit $\mu\rightarrow \infty$, and it  becomes steeper as we lower the RG scale $\mu$;  there is a crossover region around $\Lambda/\mu \sim 10^5$ where this  trend is  inverted,  and the potential starts flattening again as we lower the $RG$ scale.  This flattening continues  all the way towards the deep IR where, as we discussed  below equation (\ref{Ginf}),  
the minimum of the effective potential moves off  to infinity. 

\subsubsection{The RG-invariant gluon operator} \label{invariant}

In the previous section, we have discussed the quantum effective action for $Tr F^2$. However, notice that $TrF^2$ is not RG-invariant and there are ambiguities in the identification of the 't Hooft coupling $\lambda$ coupled to $TrF^2$ in the gravity dual when we go to the IR. An RG-invariant gluon operator is of particular interest due to its RG scale independence. One of the simplest RG-invariant operator in Yang-Mills theory is the combination  appearing in the trace identity,
\be
{T^\mu}_\mu = - {\beta(\lambda) \over 2\lambda^2} TrF^2.
\ee
Using the operator on the right hand side one can define an RG-invariant  gluon condensate, whose value is proportional to the non-perturbative scale $\Lambda$ of the theory. We will concentrate on this RG-invariant gluonic operator in the discussion below.

In the holographic theory,  the RG-invariant operator $\cal T$ associated to the bulk scalar will be   coupled to an appropriate  source  function  $J^{inv}(\lambda(x))$, rather than to $J=(-2\lambda)^{-1}$.  Then, the  vev of $\cal T$ is the functional derivative  with respect to $J^{inv}$ of the renormalized generating functional $S^{(ren)}[\gamma_{\mu\nu}, \lambda(J^{inv})]$, given in  (\ref{running}), thought as a function of $J^{inv}$.  Keeping only the  zeroth order term in derivatives,  we have
\begin{equation}\label{rginv1}
\<{\cal T}\>= \frac {\delta S^{(ren)}}{\delta J^{inv}}=- D_0\, N_c^2\xi^4 \big(\mu\, e^{- \A}\big)^4\frac {d\A} {dJ^{inv}}+O(\partial^2), \qquad \xi \equiv \left(\frac{4 M^3 \ell^3}{N_c^2}\right)^{\frac 1 4},
\end{equation}
where  $\mu=\ell^{-1}e^{A}$ is the energy scale (see equation (\ref{mu}) ).

At the zero-derivative order, the quantity $\mu\, e^{-\A}$ coincides up to a normalization constant  with the non-perturbative scale (\ref{np4}) and is itself RG invariant. Thus,  the constancy of the vev $\<{\cal T}\>$ requires $ {d\A}/ {dJ^{inv}}$ to be constant. We define: 
\begin{equation}\label{rginv2}
J^{inv}=-\A(\lambda) .
\end{equation}

With this definition \eqref{rginv2}, we see  that ${\cal T}$ coincides with the standard RG-invariant gluon condensate operator in Yang-Mills theory :
\begin{equation}\label{rginv3}
\<{\cal T}\>=  \frac {\delta S^{(ren)}}  {d (-\A)}
=\frac  {d (2\lambda)^{-1}}{d \A}\frac {\delta S^{(ren)}}  {\delta (-2\lambda)^{-1}}
=  \Big\<  -\frac {\beta(\lambda)}{2\lambda^2} \, Tr F^2\Big\>.
\end{equation}
where in the third equalities we used equation (\ref{beta}). Notice that $\<{\cal T}\>$ is positive since the $\beta$-function is negative. 

It is important to notice that the relation (\ref{rginv3}) that relates $\<\cal T\>$ and $\<-\frac {\beta(\lambda)}{2\lambda^2} \, Tr F^2\>$    is {\em independent of the identification of the bulk field $\lambda(u)$ with the Yang-Mills coupling}. As discussed in \cite{thermo3}, this identification can be established in the UV, but it could change in the IR, which introduces an extra scheme dependence in the holographic setup and makes it difficult to relate it to ordinary Yang-Mills theory. However, this ambiguity does not affect the relation (\ref{rginv3}). Indeed, suppose  we relax the indentification of $\lambda(u)$ with the field theory 't Hooft coupling $\lambda_{YM}$, and write:
\be
\lambda_{YM} = f(\lambda),  \qquad \<Tr F^2\> = {\delta\ \over \delta(-2 \lambda_{YM})^{-1}} S^{(ren)}[\lambda],
\ee
in terms of an unknown function $f(\lambda)$.
Then, we can rewrite  (\ref{rginv3}) as:
\be
\<{\cal T}\>=  {d \lambda \over d(-\A)} {d {S^{(ren)}} \over d\lambda} =- \beta(\lambda) f'(\lambda)\frac {\delta S^{(ren)}}  {\delta \lambda_{YM}} =\Big\< -{\beta(\lambda_{YM}) \over 2\lambda_{YM}^2} TrF^2\Big\>,
\ee
where in the second equality we have used the fact that $d \A/ d\lambda = 1/\beta(\lambda)$ as follows from equation (\ref{np4-1}) and $\beta(\lambda_{YM})$ indicates the beta function of the 't Hooft coupling $\lambda_{YM}$.  This shows that the holographic operator defined in (\ref{rginv1}) with $J^{inv} =- \A$ coincides with the RG-invariant gluon operator independently of the relation between the bulk field $\lambda$ and the Yang-Mills coupling. The same cannot be said for the operator dual to $\lambda^{-1}$ itself, which conicides with $Tr F^2$ in the UV but it may deviate from it in the IR.

We now procede to compute the effective potential for ${\cal T}$. The RG-invariant operator ${\cal T}$ at the zero-derivative order is computed by the functional derivative of the homogeneous part of  $S^{(ren)}$, equation (\ref{running}),    with respect to $J^{inv}$:
\begin{equation}\label{rginv4}
{\cal T}= \frac {\delta S^{(ren)}}{\delta J^{inv}}= D_0\, N_c^2\xi^4 \big(\mu e^{- \A}\big)^4,
\end{equation}
where we have used equation \eqref{Z0}  in the second equality. 

If we recall the definition of the non-perturbative scale $\Lambda$  (\ref{np4}), with our choice of the normalization, we see that the right hand side of (\ref{rginv4}) is nothing but $D_0N_c^2\Lambda^4$.
It is convenient to fix  the scheme such that $D_0=1$, so that the vacuum expectation value of ${\cal T}$ is simply given by:
\be\label{rginv5}
\<{\cal T}\> = N_c^2 \Lambda^4, \qquad D_0 =1.
\ee

The  effective potential for ${\cal T}$, i.e.  the zero-derivative part  of the  effective action, is given by: 
\be
{\cal V}[{\cal T}] = \left[ \big(J^{inv}-\bar J^{inv}\big) {\cal T}-\frac {N_c^2\xi^4}4\mu^4Z_{0}\right], \qquad \xi \equiv \left(\frac{4 M^3 \ell^3}{N_c^2}\right)^{\frac 1 4},
\ee
where the last term is the  homogeneous part of $S^{(ren)}$,  given  in equation \eqref{Z0}.

Using the full RG invariant coupling
\be\label{rginv7}
J^{inv}=-\A\left[\frac {\cal T} {\mu^4}\right]=\frac 1 4 \ln \left(\frac {\cal T}{N_c^2 \xi^4 \mu^4}\right)
\ee
as a function of the vev ${\cal T}$ calculated from \eqref{rginv4} and the background RG invariant coupling
\be\label{rginv8}
\bar J^{inv}=-\bar \A\left[\frac {\Lambda} {\mu}\right]=\frac 1 4 \ln \left(\frac {\Lambda^4}{\xi^4\mu^4}\right)
\ee
computed by inverting the definition of the non-perturbative scale
\be
\Lambda= \xi\,\mu \,e^{-\bar\A},
\ee
the effective potential reads
\be \label{rginv6}
{\cal V}[{\cal T}] =\frac {\cal T} 4\left(\ln\frac {\cal T}{N_c^2\Lambda^4}-1\right)
\ee

By construction, we have:
\be
\frac \delta{\delta{\cal T}}{\cal V}({\cal T}) = 0 \quad \Leftrightarrow \quad {\cal T}_{\min} = N_c^2 \Lambda^4,
\ee
which is consistent with \eqref{rginv4}.

From the final result of \eqref{rginv6}, we can see that the effective potential is model-independent and it has a universal form for arbitrary bulk scalar potential $V(\phi)$. The RG scale $\mu$ dependence in the effective potential is cancelled as well, so the form of $\Gamma^{(0)}[{\cal T}] $ is RG invariant, which is natural for an RG invariant operator.

\begin{figure}[h!]
\begin{center}
\includegraphics[width=8cm]{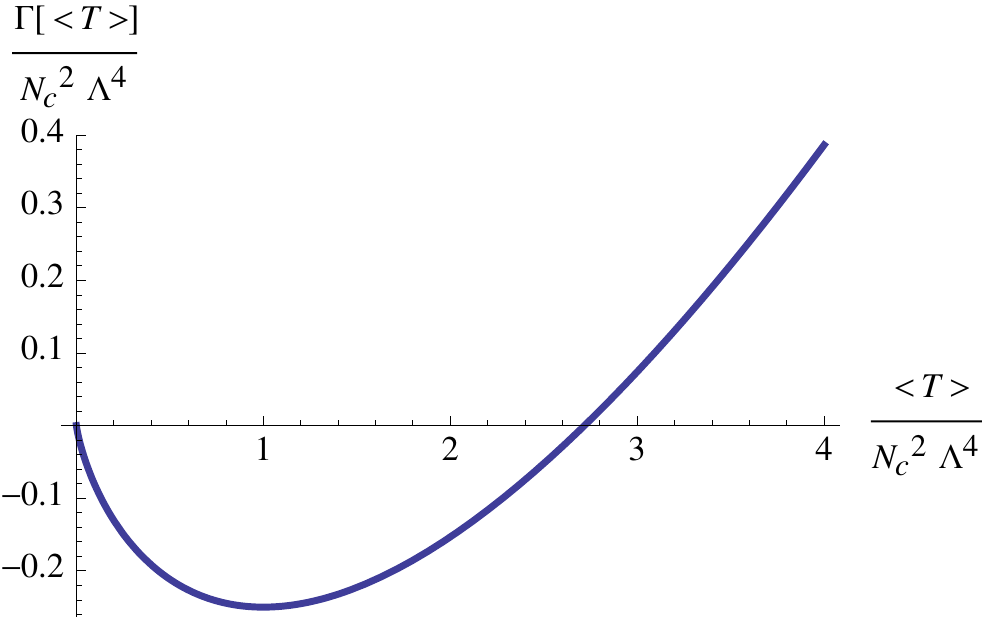}
\caption{The model-independent effective potential $\Gamma^{(0)}[\<{\cal T}\>]$ of the RG-invariant operator $\<{\cal T}\>$, whose analytic form is in (3.80). 
The scheme is chosen in such a way that the minimum is located at $\<{\cal T}\>=N_c^2\Lambda^4$, where $\Lambda$ is the non-perturbative energy scale defined in Section 2.3.
}
\label{pot-inv}
\end{center}
\end{figure}

As before, going to second order in space-time derivatives and substituting \eqref{rginv7} into the general formula (\ref{1PI}), one can derive the two-derivative terms in the 1PI effective action, from which one is able to find out the canonically normalized operator ${\cal T}^{(c)}$ using the kinetic term and to write the 1PI action in terms of the canonically normalized operator ${\cal T}^{(c)}$.

Here we will present the analytic results of the RG-invariant operator for the UV and IR, where one can expand the superpotential for small and large $\lambda$ respectively.

In the UV, the RG invariant operator coincides with $Tr F^2$ up to a numerical factor $b_0$ since $\A\sim \frac 1{b_0\lambda}$ from the UV expansion \eqref{A-UV} and thus ${\cal T}\sim \frac {b_0}2{\cal G}$. We can obtain the effective action of $\<{\cal T}\>$ by simply substituting $\frac {b_0}2{\cal G}$ by ${\cal T}$, so the result is
\be
\Gamma_{UV}[{\cal T},\Lambda] =\int \left[ \frac {  {\cal T}}4 \left( \ln\frac{{ \cal T}}{N_c^2\Lambda^4}  -1 \right)
+\frac {b_0^{ 2}\,N_c\,\xi^2}{64} {\cal T}^{-\frac 3 2}\frac 1 2 (\partial{\cal T})^2\right], \quad \xi \equiv \left(\frac{4 M^3 \ell^3}{N_c^2}\right)^{\frac 1 4},
\ee

From the kinetic term, we define the canonically normalized operator ${\cal T}^{(c)}$ as
\be
{\cal T}^{(c)}=\frac {b_0\,N_c^{\frac 1 2}\,\xi} {2} {\cal T}^{\frac {1}{4}}={\cal G}^{(c)},
\ee

and the corresponding effective action has the same form as \eqref{EP-G-CN}
\be\label{rginv7-ii}
\Gamma_{UV}[{\cal T}^{(c)},\Lambda]=\int \left[ \frac {4{{\cal T}_{(c)}}^4}{\xi^4 b_0^{4}N_c^2}\Big( \ln \frac {{16{\cal T}_{(c)}}^4}{\xi^4b_0^4N_c^4\Lambda^4} -1\Big)
+ \frac 1 2\eta^{\mu\nu}\partial_\mu {{\cal T}^{(c)}}\partial_\nu {{\cal T}^{(c)}} \right].
\ee

The identification of the RG invariant operator $\cal T$ with the YM operator $\cal G$ in the UV explains the fact that the $\Gamma_{UV}[{\cal G}]$ is also RG scale invariant.

In the IR, the effective action of the RG invariant operator $\cal T$ is different from that of the YM operator $\cal G$. To compute the effective action, we first write the IR renormalized generating functional in a convenient form
\be
S^{(ren)}=M^{3}\ell^{3}\int d^4 x\,\left[ \mu^4e^{-4\A}+D_3\mu^{2}  \frac 1 2 \eta^{\mu\nu}\big(\partial_\mu e^{-\A}\big)\big(\partial_\nu e^{-\A}\big)\right],
\ee
where 
\begin{equation}
 D_3=\ell^{2} G_0^{(1)}(IR)\frac  {16} {3Q^2}, \qquad (Q=2/3\; \text{for Yang-Mills duals})
\end{equation}
and we have neglected the subleading terms in the $\lambda\rightarrow \infty$ limit. Then we can use the equations \eqref{rginv4}, \eqref{rginv7}, \eqref{rginv8} and \eqref{1PI} to derive the effective action $\Gamma[\cal T]$
\be
\Gamma_{IR}[{\cal T},\Lambda] =\int d^4 x \,\left[\frac {\cal T} 4\left(\ln\frac {\cal T}{N_c^2\Lambda^4}-1\right)
+\frac {D_3N_c\xi^2} 4  \frac 1 2 \big(\partial {\cal T}^{\frac 1 4}\big)^2\right], \quad \xi \equiv \left(\frac{4 M^3 \ell^3}{N_c^2}\right)^{\frac 1 4}.
\ee
From the explicit form of the kinetic term, we determine the canonically normalized operator to be
\be
{\cal T}^{(c)}=\frac {N_c^{\frac 1 2}\xi\,D_3^{\frac 1 2}} {2} {\cal T}^{\frac 1 4}.
\ee
Finially, we derive the IR-limit of the effective action of the canonically normalized RG-invariant operator ${\cal T}^{(c)}$
\be \label{rginv9}
\Gamma_{IR}[{\cal T}^{(c)},\Lambda] =\int d^4 x \,\left[\frac {4{\cal T}_{(c)}^4} {N_c^2\xi^4 D_3^2}\left(\ln\frac {16\,{\cal T}_{(c)}^4} {N_c^4\xi^4 D_3^2\Lambda^4}-1\right)
+\frac 1 2 \eta^{\mu\nu}\partial_\mu {\cal T}^{(c)}\partial_\nu {\cal T}^{(c)}\right].
\ee
Looking at  equations (\ref{rginv7-ii}) and  (\ref{rginv9}), notice that the UV and IR effective potentials of the canonically normalized vev ${\cal T}^{(c)}$ have the same functional form:
\be
\Gamma^{(0)}[{\cal T}^{(c)},\Lambda]= \int \frac {{\cal T}_{(c)}^4}{\,\alpha} \left(\ln \frac {4{\cal T}_{(c)}^4}{\alpha\,\Lambda^4}-1\right),\quad \alpha_{UV}=\frac {N_c^2\xi^4b_0^{4}} 4,\quad \alpha_{IR}=\frac {N_c^2\xi^4D_3^2} 4,
\ee
but the minima are different due to the change of $\alpha$.

\subsubsection{The bare gluoninc operators}
In this section  we will compute the holographic bare, regularized effective action. This is done by Legendre transformation of the bare action for the sources with an explicit UV boundary cut-off. We first calculate the effective action for the gluonic operator ${\cal G}=Tr F^4$,  then the one for the RG-invariant operator ${\cal T}=  (-\beta_\lambda) \, Tr F^2 /{(2\lambda^2)}$.

Using the notation introduced in Subsection \ref{scheme}, the zero-derivative term in the bare action according to \eqref{Sreg2} reads
\begin{equation}
S^{(bare)}[e^{A}=\epsilon^{-1} \ell,\lambda]=M^{3}\ell^4\int d^4 x\,\epsilon^{-4}\,W(\lambda).
\end{equation}

To compute the vev of bare gluon operator $Tr F^2$, as in Subection \label{secTrF2} we take the functional derivative of the regularized action $S^{(bare)}$ with respect to the source $J=(-2 \lambda)^{-1}$, 
\be\label{G-reg-UV}
\< {\cal G}_\epsilon\>=\frac {\delta S^{(bare)}} {\delta (-2\lambda)^{-1}}={N_c^2\,\xi^4 \,\ell \over {2\epsilon^4}}  \lambda^2\frac{dW}{d\lambda}.
\ee

To carry out the explicit computation of the effective action, we need to make a natural assumption that the background value of the coupling  at the cut-off is very small,  
\be\label{small-lambda}
\lambda_\epsilon\ll 1,
\ee
so the perturbative expansions around the UV fixed point is possible. 
As explained at the end of Subsection  \ref{scheme},  the assumption \eqref{small-lambda} implies that the UV cut-off scale $\Lambda_{UV}$ is much large than the non-perturbative scale $\Lambda$, 
\be\label{small-lambda-ii}
 \frac \Lambda{\Lambda_{UV}}=\epsilon\, \Lambda\ll 1.
\ee

The equivalence  \eqref{small-lambda} and (\ref{small-lambda-ii})  can be understood from the definition of the non-perturbative scale \eqref{np6},
\be\label{Lambda-reg-UV}
\Lambda= \xi\,\Lambda_{UV} \,e^{-\A(\lambda_\epsilon)},
\ee
where the physical scale $\mu$ replaced  by the lattice cut-off $\epsilon=\Lambda_{UV}^{-1}$.

Under the small $\lambda_\epsilon$ assumption, the vev of ${\cal G}_\epsilon$ at the leading order becomes
\be
{\cal G}_\epsilon=\frac {11N_c^2\xi^4}{18\pi^2}
{\lambda^2 \over  \epsilon^{4}}
\ee
from which one can invert the relation between $\<{\cal G}_\epsilon(x)\>$ and $\lambda$ to obtain the 't Hooft coupling as a function of the vev
\begin{equation}
\lambda({\cal G}_\epsilon)=\sqrt{\frac{18\pi^2} {11N_c^2\xi^4}
\epsilon^{4}\,{\cal G}_\epsilon}\, .
\end{equation}

The background source at the cut-off at the leading order of small $\lambda_\epsilon$ expansion can be calculated as well
\be
\frac 1 {\lambda_\epsilon}=-b_0\ln \left ( \frac {\Lambda}{N_c^{\frac 1 2}\xi\,\Lambda_{UV}} \right),
\ee

from which we can compute the source
\be
\tilde J(x)=J(x)-J_\epsilon=\frac 1 {2 \lambda_\epsilon}-\frac 1 {2\lambda(x)}.
\ee

Then we calculate the effective action by Legendre transform of the bare action \eqref{Sreg} with respect to the source $\tilde J$
\begin{align}
\Gamma_\epsilon[{\cal G}_\epsilon,\Lambda,\epsilon]=&\int d^dx\,{\cal G}_\epsilon \left[\frac 1 {2 \lambda_\epsilon}-\frac 1 {2\lambda(x)}\right]-S^{(bare)}\nonumber\\
=&\int \,\frac {{\cal G}_\epsilon }{2\lambda_\epsilon}
-M^{3}\ell^4\epsilon^{-4}\int\,\Big[W(\lambda)+\lambda \frac {d\,W(\lambda)}{d\, \lambda}\Big]\nonumber\\
&-M^{3}\ell^{2}\epsilon^{-2}\int \frac {W\partial_\lambda U}{\partial_\lambda W}  \frac 4 3(\partial\log \lambda)^2,\nonumber\\
=&M^3\ell^3\int \,\left\{\epsilon^{-4}
\left[-6-\frac{11\sqrt 2} {3\pi N_c\xi^2}\big(\epsilon^4\,{\cal G}_\epsilon\big)^{\frac 1 2}\right]
+\,\epsilon^{-2}\frac 1 6 {\cal G}_\epsilon^{-2} (\partial {\cal G}_\epsilon)^2\right\}
\end{align}
where higher order terms of ${\cal G}_\epsilon$ are neglected.

The effective potential can be put in a form which makes manifest the violation of scale invariance:   
\be
{\cal V}_\epsilon[\epsilon, {\cal G}_\epsilon] =\epsilon^{-4}\,v[x]+ \frac {{\cal G}_\epsilon}{2\lambda_\epsilon},\qquad x\equiv\epsilon^4\, {\cal G}_\epsilon,
\ee
both $x$ and $v[x]$ are invariant under a change of cutoff $\epsilon$ and thus the first term on the right hand side is scale invariant. However, the background term introduces scale violation into the effective potential. The background source $\lambda_\epsilon=\lambda_\epsilon(\epsilon^4\Lambda)$
is an explicit function of the cutoff due to the constancy of the non-perturbative scale. For example, at the leading order of small $\lambda_\epsilon$ expansion
\be
\frac{{\cal G}_\epsilon}{\lambda_\epsilon}=\epsilon^{-4} (\epsilon^4 {\cal G}_\epsilon)\lambda_\epsilon^{-1}\sim \epsilon^{-4}\, x\, \log(\epsilon\, \Lambda) ,\qquad x=\epsilon^4\, {\cal G}_\epsilon,
\ee
so the effective potential have a logarithmic scaling violation.

The effective action of RG-invariant operator ${\cal T}_\epsilon$ has the same form since in the UV region the RG-invariant coupling is the same as the YM operator up to a numerical factor
\be
J^{inv}=-\frac 1{b_0\lambda}=\frac 2 {b_0}\, J\,\Rightarrow\, {\cal T}_\epsilon=\frac {b_0} 2\,  {\cal G}_\epsilon.
\ee
 
From the kinetic term, one can determine the canonically normalized operator to be
\be
{\cal G}_\epsilon^{(c)}={\cal T}_\epsilon^{(c)}=M^{\frac 3 2} \ell \,\epsilon^{-1}\left[\ell^{\frac 1 2}+\int_{\phi_\epsilon} ^\phi d\psi \left(\frac {W\partial_\lambda U}{\partial_\lambda W}\right)^{\frac 1 2}\right],
\ee
which is the same for different operators due to the same form of kinetic terms when written as a function of $\phi$. The expression for the zero-derivative term ( the potential term) depends on the specific definitions of its operators, but they coincide in the small $\lambda_\epsilon$ limit. We have fixed the integration constant in such a way that
the value of the canonically normalized operator at the minimum of the effective potential is
\be
\epsilon\,(M\ell)^{-\frac 3 2}  [\cO^{(c)}_\epsilon]_{\min}=1.
\ee

\begin{figure}[h!]
\begin{center}
\includegraphics[width=10cm]{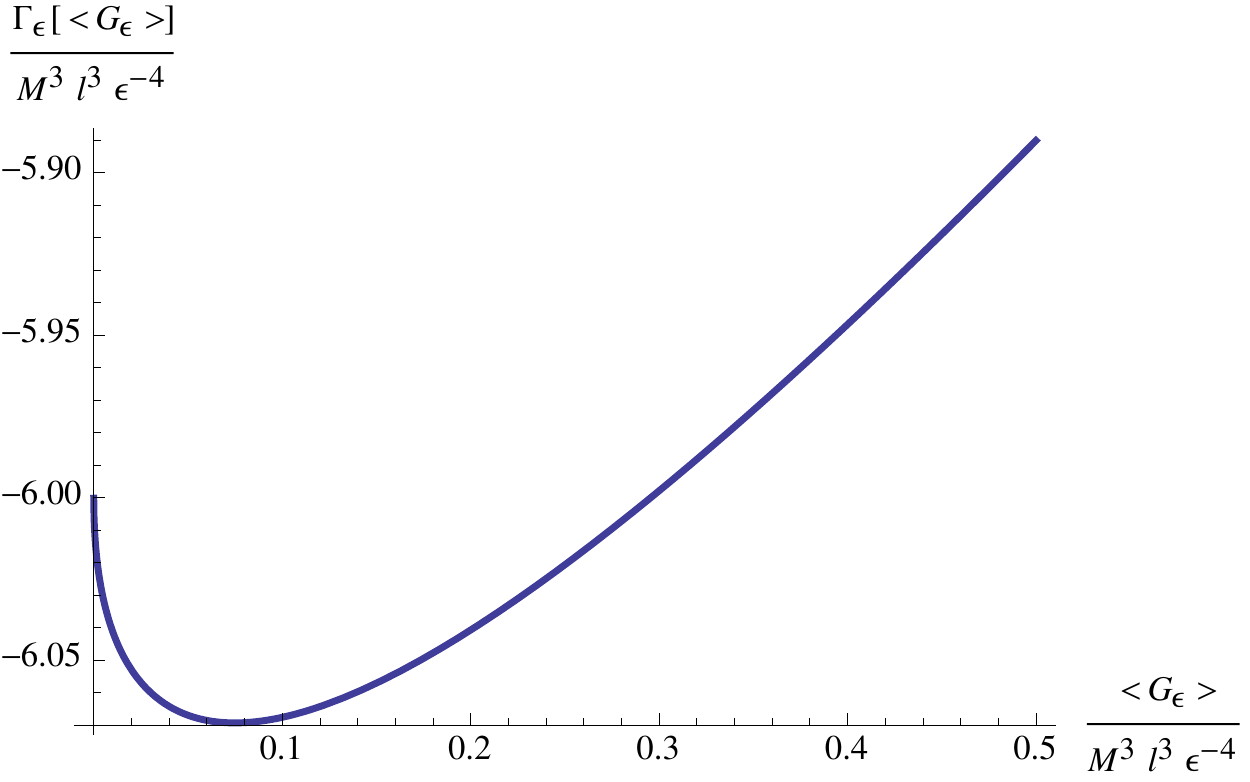}
\caption{The bare effective potential of ${\cal G}$. The background source is chosen to be $\lambda_\epsilon=0.5$, which can be translated into the value of the cut-off energy $\Lambda_{UV}\sim \Lambda\times 10^{20}$. The potential has a universal small $\cal G$ limit $\Gamma[{\cal G}\rightarrow 0]=-M^3\ell^4\epsilon^{-4}W[\lambda\rightarrow 0]=-6M^3\ell^3\epsilon^{-4}$. For large $\cal G$, the potential is dominated by the linear background term with a slope $(2\lambda_\epsilon)^{-1}\sim 1$.}
\label{sim-pot-bare}
\end{center}
\end{figure}

As an illustration, we compute the effective potential of ${\cal G}_\epsilon$ using the simple potential \eqref{sim-pot}, and we present the result in Figure \ref{sim-pot-bare}.

Now let us see explicitly how the renormalized v.e.v. $\cal T$ is obtained by subtracting the divergence pieces of the bare v.e.v. ${\cal T}_\epsilon$ by counter-terms. In terms of the cutoff $\epsilon$, the bare vev at the minimum of the effective potential is
\be
{\cal T}_\epsilon=M^{3}\ell^4\epsilon^{-4}\,\frac {d W}{d(-A)}=\frac 4 {3} N_c^2\xi^4\epsilon^{-4}\,\left [\,\ln  (\epsilon\,\Lambda) \right]^{-2}+...+CN_c^2\Lambda^4+...
\ee
where the finite terms start from $C_\epsilon\,\Lambda^4$ and $C_\epsilon$ is the integration constant in the regular superpotential solution.
The subleading divergent terms and the subleading finite terms are not written explicitly.

Since the counter-terms in the renormalized action have the same form as the bare Lagrangian but with different integration constants, the universal divergent terms in the bare v.e.v. will be subtracted. Of the remining  terms, the ones that remain finite as we move the cut-off to infinity are:
\be
{\cal T}=\lim_{\epsilon\rightarrow 0} ({\cal T}_\epsilon-{\cal T}_C)=\lim_{\epsilon\rightarrow 0} \left[(C -C_{ct})N_c^2\Lambda^4+O(\epsilon)\right]=(C -C_{ct})N_c^2\Lambda^4,
\ee
where $C_{ct}$ is the integration constant that governs the subleading part of the counterterm superpotential.  The remaining subleading terms vanish as $\epsilon \to 0$. The coefficient $D_0$ in \eqref{Z0} is $D_0=C-C^{ct}$.

 \section{Effective potentials in the realistic IHQCD model}

In this section we give a concrete example based on the Improved Holographic QCD model \cite{ihqcd}.
Our starting point is the full bulk potential of IHQCD:
\begin{equation}
V(\lambda)=\frac {12}{\ell^2}\left\{1+V_0\lambda+V_1\lambda^{\frac 4 3}\left[\log(1+V_2\lambda^{\frac 4 3}+V_3\lambda^2)\right ]^{\frac 1 2}\right\}
\end{equation}
where $V_0=\frac {11} {27\pi^2}$, $V_1=14$, $V_2=\left(\frac {11}{24\pi^2}\right)^4\left(\frac {1836}{121}\right)^2$ and $V_3=170$. The AdS length $\ell$ sets the unit of energy and does not appear in the dimensionless physical quantities. $V_0$ and $V_2$ are determined by the 2-loop $\beta$-function of QCD. $V_3$ and $V_4$ are phenomenological parameters corresponding to the best fit to the thermodynamic functions. This bulk potential interpolates between the UV and the IR asymptotic behaviors (\ref{V-UV}) and (\ref{V-IR}) with $Q=2/3, P=1/2$. 

The solution for superpotential $W(\phi)$ is obtained by solving numerically equation (\ref {superpotential-eq}) and imposing IR regularity, i.e. the condition (\ref{W-IR}).\\

It is interesting to estimate the non-perturbative scale $\Lambda$, which we defined as (\ref{np6}), that   corresponds  to the fisical choice of units that matches real-life Yang-Mills theory. This can be computed, for example,   by fixing the bulk solution and the asymptotic AdS scale $\ell$ in such a way that the lowest $0^{++}$ glueball mass  coincides  with the value calculated on the lattice\footnote{Any other dimensionful physical observable, e.g. the deconfinement temperature $T_c$,  would have done the job.}, $m_{0^{++}} = 1710~MeV$ \cite{chenetal}. The resulting non-perturbative scale is
\be\label{lambdaYM}
\Lambda = 191~MeV
\ee
and it agrees with what is generally taken to be the  scale of non-perturbative Yang-Mills theory, 
 confirming  that our definition of the non-perturbative scale can be matched concretely to the field theory result.   

\begin{figure}[h!]
\begin{center}
\includegraphics[width=12cm]{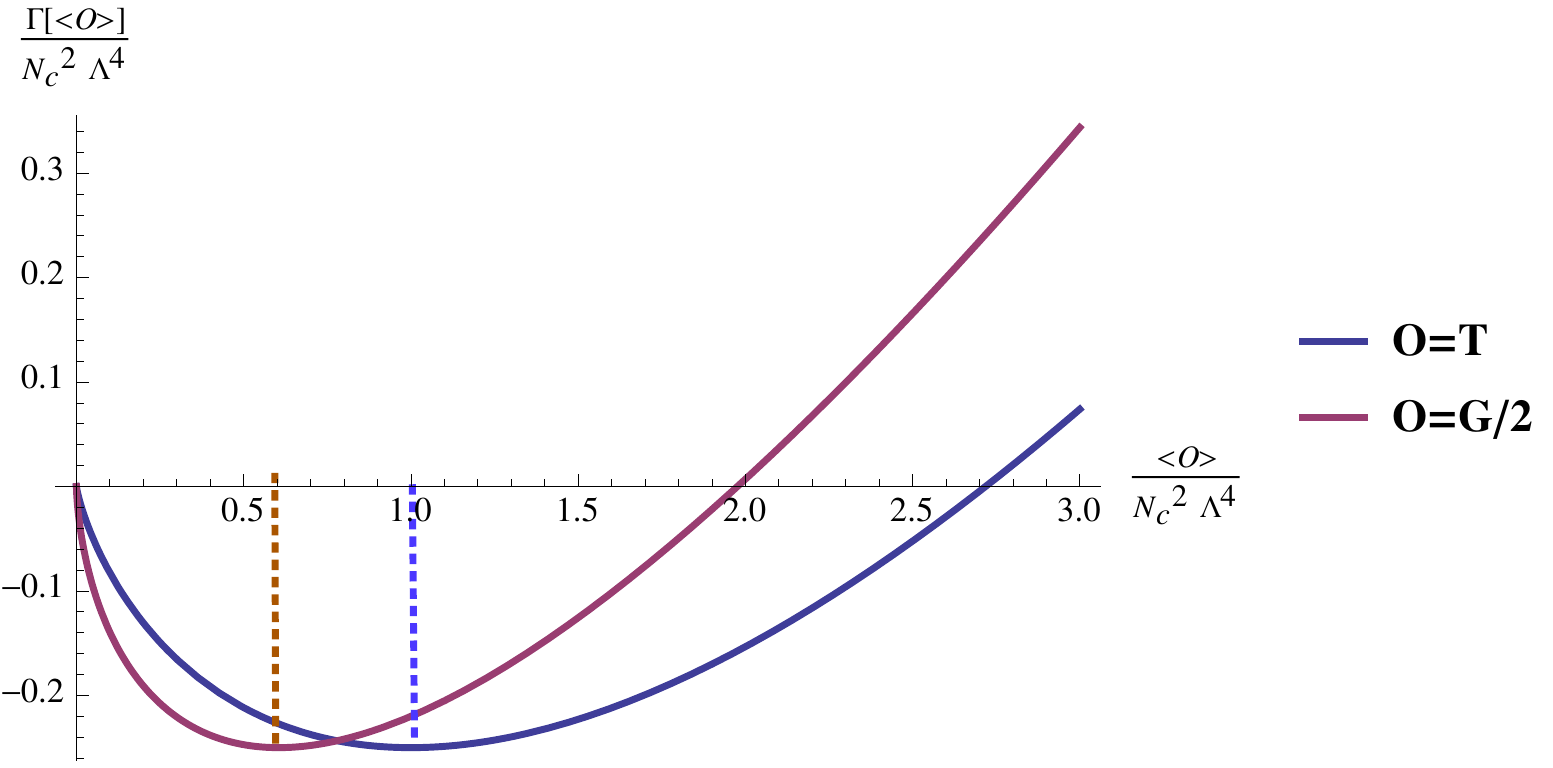}
\caption{The two curves represent the effective potentials $\Gamma$, in units of $\Lambda^4 N_c^2$,  for the RG-invariant operator ${\cal T}=  -\beta_\lambda/(4{\lambda^2}) \, Tr F^2$ (blue) and ${\cal G}=Tr F^2$ (red). For the latter operator, the horizontal axis is rescaled by a factor of 1/4 to achieve better graphical clarity.   The minimum of $\Gamma[{\cal T}]$ is at ${\cal T}_{\min}=N_c^2\Lambda^4$, as indicated in the analytic form (3.80), while the minimum of $\Gamma[{\cal G}]$ is at ${\cal G}_{\min}\sim 1.2N_c^2\Lambda^4$. } \label{pot-tr-inv}
\end{center}
\end{figure}

\begin{figure}[h!]
\begin{center}
\includegraphics[width=12cm]{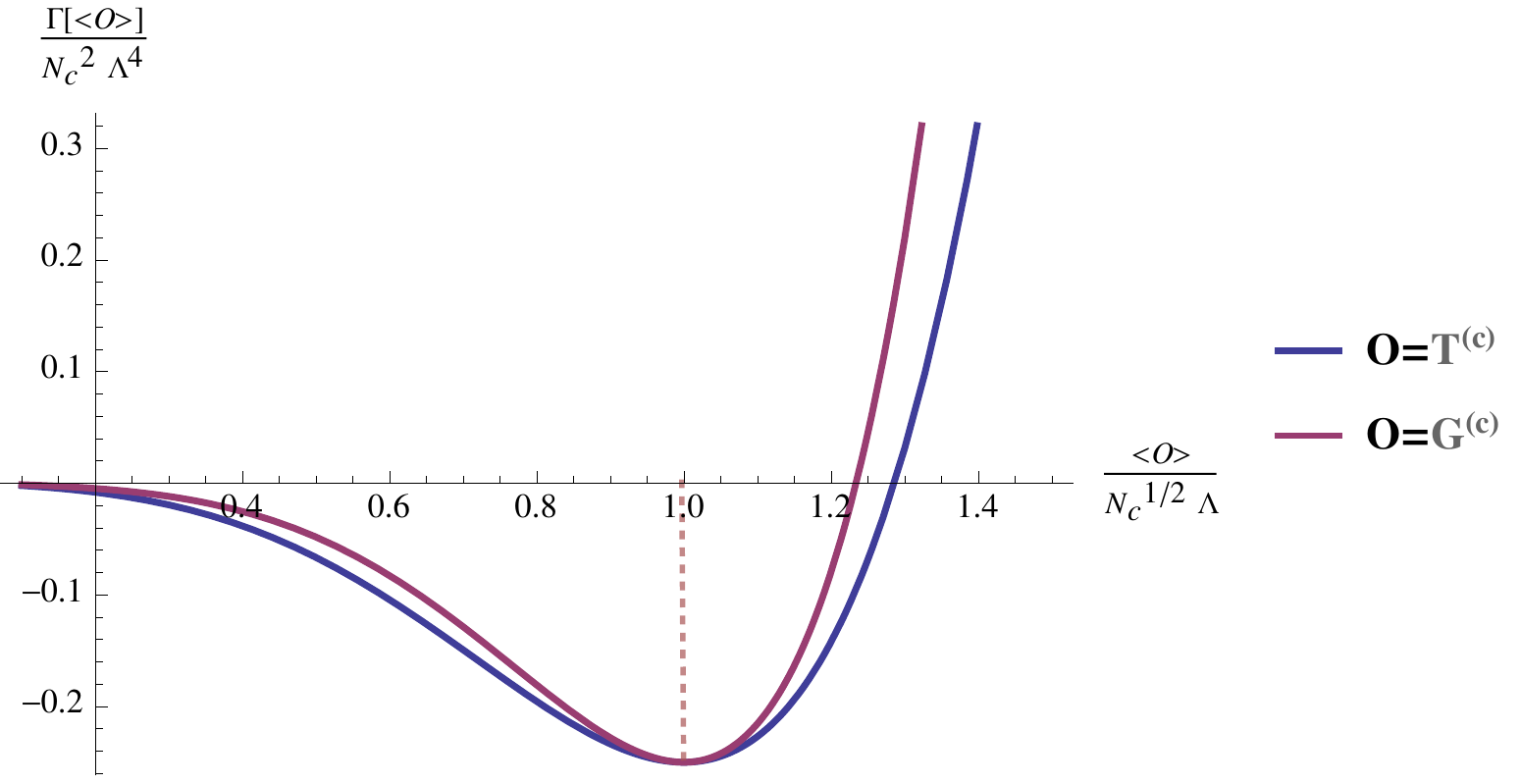}
\caption{The effective potentials of canonically normalized operators. The blue curve is the effective potential of $ -\beta/(4\lambda^2) \, Tr F^2$ and the red one denotes the effective potential of $Tr\,F^2$. The minima of the two potentials coincides because the kinetic terms as a function of $\phi$ are the same. The value of the vev at the minimum is scheme-dependent. Here it is located at ${\cal T}^{(c)}={\cal G}^{(c)}=N_c^{\frac 1 2}\Lambda$ due to our choice of scheme $D_0=1,\, D_1=0,\, D_2=1.69$ .} \label{pot-ca-tr-inv}
\end{center}
\end{figure}

Using the numerical superpotential solution, one can compute the zero-derivative term and the coefficients of the two-derivative terms in the renormalized generating functional. Then we compute the effective potential of the renormalized composite gluon operator ${\cal G}=Tr F^2$ and the renormalized RG-invariant operator ${\cal T}= -{\beta(\lambda) } TrF^2/ 2\lambda^2$. The results are displayed  in Figure \ref{pot-tr-inv} and Figure \ref{pot-ca-tr-inv}. In our numerical calculation, we have fixed  the energy scale for $\cal G$ such that  the background sources take simple values:
\be
\phi(\mu)=0\Leftrightarrow \lambda(\mu)=1\Leftrightarrow \A(\mu)=0.
\ee
The scheme dependent coefficients $D_0$ and $D_1$ have been chosen as in equation (\ref{Dis}), but $D_2$ is chosen differently to set the vev of canonically normalized operator at the minimum to be $N_c^{1/2}\Lambda$.  
Of the curves shown in figures \ref{pot-tr-inv} and \ref{pot-ca-tr-inv}, only the blue one in  figure \ref{pot-tr-inv} is independent of the choice of scale, as it corresponds to the effective potential for the un-normalized RG-invariant operator. Changing the choice of the reference  scale affects the\
other curves in a way similar to that shown in figure \ref{simp-pot-UV-IR}. 
Notice that the canonically normalized  operators ${\cal T}^{c}$ and ${\cal G}^{c}$ have the same v.e.v. This is because, as explained below eqaution (\ref{vev-ii}), both v.e.v.'s are given in terms of the same function of the scalar field $\phi$. 
 
\section{Conclusion}
In this work we have shown, in a phenomenological gravity dual in five dimensions,  how to compute the effective action, up to two derivatives,  for the lowest dimension, scalar single trace operator in Yang-Mills theory: the dimension-four scalar  glueball operator.   

 We have found for the RG-invariant  glueball operator a  universal, simple analytic  form for the potential, given in equation (\ref{rginv6}),  which nicely incorporates  the conformal anomaly. Notice that this form is not restricted to QCD-like theories, but to any holographic model driven by a single scalar. It would be interesting to extend this  and the other calculations in this paper to the multi-field case and see whether this simple structure persists.  

Switching  to canonical normalization the resulting potential ceases to be  universal, because the effective kinetic term does depend on the specific bulk theory.

Beside presenting the calculational framework and providing model-independent results, we have also analyzed in detail the effective potentials arising 
in specific models which were proposed as realistic phenomenological gravity duals of Yang-Mills theory. These models are known to reproduce very accurately the  static thermal  properties of finite-temperature Yang-Mills theory \cite{ihqcd}, and  it  would be interesting to investigate whether this agreement extends to other static quantities like the gluonic effective potential. 
The main obstacle here is the fact that a reliable lattice computation of this quantity is hard to achieve, due to the amount of short-distance noise in this channel.  However, the only source of scheme dependence is in the overall scale of the potential, not its shape: thus,  if one could manage to subtract the UV contribution to the vacuum energy, the shape of the remaining effective potential should be completely fixed, and comparison  with the holographic results presented here should be possible. 

The  effective actions we constructed  encode the full quantum  dynamics  in cases when it is governed by  a single operator of the theory, and mixing with other single-trace operators is weak. This is not necessarily the case for the true Yang-Mills theory, but it can be used  as a first approximation, which can be in principle be checked, for example on the lattice. 

Although it has the standard form of Lorentz-invariant kinetic plus potential terms,  the effective action  we obtained must {\em not} be interpreted as the action for a field describing physical particle excitations. Thus, small oscillations around the vacuum do not describe modes  that have a  direct particle interpretation\footnote{In fact, it would be wrong to think of the effective action as the starting point  for quantization.} .

Rather, the effective action  encodes the collective behavior of the full tower of physical particle modes (in this case, the tower of $0^{++}$ scalar glueballs) associated to the gauge-invariant operator in question. Our holographic  models, like  Yang-Mills theory, have an infinite discrete spectrum of states, whose masses $m_n$  are set by the non-perturbative scale $\Lambda$. These states are the eigenmodes of the linearized scalar  bulk fluctuations, and they appear as poles in the two-point function of the dual operator ${\cal O}$:
\be\label{conc1}
\< {\cal O}(p) {\cal O}(-p)\> = \sum_{n = 0}^{\infty} {f_n^2 \over p^2 + m^2_n} + \text{Analytic in $p^2$} 
\ee
Generically all low-lying states have comparable masses and decay constants, 
Thus,  unlike the case of  a (quasi)-free field,   the dynamics  around the vacuum cannot be associated to any one particular physical particle. 

To illustrate this point further, let us consider ${\cal O}(x)$  as a small perturbation around the vacuum $\<\cal O\>$, and let us expand the quantum effective action, of the general form (\ref{intro1}),   to quadratic order in ${\cal O}(x)$ as well as second order in momenta: 
\be\label{conc2}
\Gamma[{\cal O}] = \Gamma_0 + {1\over 2} \int {d^4p \over (2\pi)^4} \Gamma_2(p){\cal O}(p){\cal O}(-p).
\ee
In the above expression,  $\Gamma_0$ is the vacuum energy, and $\Gamma_2(p)$ is the ``self-energy,''
\be
\Gamma_0 = {\cal V}({\cal O}), \qquad \Gamma_2(p)={\cal V}''(\<{\cal O}\>) + G(\<{\cal O}\>) p^2 + O(p^4)
\ee
which we have written up to quadratic order  in momenta, in terms of the functions ${\cal V}$ and $G$ parametrizing (\ref{intro1}), here evaluated on the vacuum $\<{\cal O}\>$. 
    
Since $\Gamma[{\cal O}]$ is the Legendre transform of the generating functional  of connected correlators, we have the standard relation between the two-point function (\ref{conc1}) and the self-energy,
\be\label{conc3}
\Gamma_2(p) = {1 \over \<O(p)O(-p)\>}. 
\ee 
Thus, using (\ref{conc1}) for the right hand side and  expanding it  to second order in momenta we can find relations, in the form of sum rules, between the spectral data $m_n,f_n$ and the ``mass'' and ``kinetic'' term appearing in the effective action:
\be\label{conc4}
{\cal V}''(\<{\cal O}\>) = {1\over C_0 + \sum_n (f_n/m_n)^2} , \qquad {\cal G}(\<{\cal O}\>) =  {C_2 + \sum_n (f_n/m_n^2)^2 \over \left({\cal V}''(\<{\cal O}\>)\right)^2 }.
\ee
In these expressions, $C_0$ and $C_2$ are the  coefficients of the constant and $p^2$ term in expansion of the analytic part of the two-point function (\ref{conc1}). These correspond to contact terms which have UV divergences and are subject to renormalization. Thus, their finite part is scheme dependent, and this must  match the scheme dependence of the coefficients of the effective action on the left hand side of (\ref{conc4}). 

It would be interesting to check these sum rules explicitly in models with a known spectrum.  We leave a more detailed  investigation of this problem,  and of the precise  way scheme dependence enters into the sum rules, for future work. 
  
From equation  (\ref{conc4}) it is clear that the second derivative of the potential cannot be associated to any one particle mass, but it is given by a collective effect of all physical particle excitations. This is to be contrasted with the case of the potential in  a weakly coupled field theory.  

One interesting  exception is when one of the modes is much lighter than the others, i.e. when the theory has a light dilaton-like particle excitation: in this case the first term in the sums dominates, the dynamics is dominated by the light mode, and the field $O$ behaves approximately like a free field whose associated particle is the light mode. In these special cases, the effective action we have constructed can be thought of as the effective action for the composite light modes, in the spirit of \cite{megias}.  However, in holographic models this situation seems to be non-generic, and requiring fine tuning \cite{piai}.

In the generic case, on the other hand, assuming ${\cal O}$ is the operator driving the dynamics, the effective actions computed in this paper will describe the vacuum structure and evolution purely in classical terms, as 
a collective behavior of ${\cal O}$ seen as a classical field, and there will be no narrow-width particle-like excitations.  The framework we have developed thus
offers an intuitive  Lagrangian tool to describe the dynamics of condensates which are not necessarily associated to particles, and can find many uses in 
holographic phenomenology.

\section{Acknowledgements}\label{ACKNOWL}

We would like to thank M. Panero and A. Patella for discussion. 
This work was supported in part by European Union's Seventh Framework Programme under grant agreements (FP7-REGPOT-2012-2013-1) no 316165,
PIF-GA-2011-300984, the EU program ``Thales'' MIS 375734, by the European Commission under the ERC Advanced Grant BSMOXFORD 228169 and was also co-financed by the European Union (European Social Fund, ESF) and Greek national funds through the Operational Program ``Education and Lifelong Learning'' of the National Strategic Reference Framework (NSRF) under ``Funding of proposals that have received a positive evaluation in the 3rd and 4th Call of ERC Grant Schemes''. We also thank the ESF network Holograv for partial support.


\newpage
\appendix
\renewcommand{\theequation}{\thesection.\arabic{equation}}
\addcontentsline{toc}{section}{Appendix}
\section*{Appendix}

\section{The effective potential for  $\<Tr F^2\>$}

\subsection{1PI effective action in the UV}
In the UV region, the function $\cal A(\phi$) in the renormalized action can be expressed in terms of $\lambda$ as
\begin{equation}\label{A-UV}
\A(\lambda)=-\frac 1 {6}\int^\phi d \tilde \phi \frac W {W'} =\frac 1 {b_0  \lambda}+\A^\ast+a\ln\lambda+O( \lambda),\qquad \lambda=e^{\sqrt{\frac 3 8}\phi},
\end{equation}
where the UV solution (\ref{W-UV}) of the superpotential has been used. $a$ is determined by the two loop $\beta$-function and its precise value is irrelevant to the discussion below. $\lambda$ coincides with the 't Hooft coupling as $\lambda\rightarrow 0$.

The renormalized generating functional \eqref{running} in terms of the 't Hooft couping $\lambda$ is:
 \begin{align}
S^{(ren)}=&M^{3}\ell^{-1}\int d^4 x\,\Big\{\big(\mu\ell\big)^4D_0 e^{ -\frac 4{b_0\lambda}-4 \A^\ast}\lambda^{-4a}\big[1+O(\lambda)\big]\nonumber\\
&\qquad\qquad\qquad+\big(\mu\ell\big)^2  \tilde D_2 \ell^2e^{-\frac 2 {b_0\lambda}-2\A^\ast} \lambda^{-2a-4}\big[1+O(\lambda)\big]\frac 1 2 \eta^{\mu\nu}\partial_\mu\lambda\partial_\nu\lambda\Big\},\label{ihqcdaction-UV}
\end{align}
where the constant $\tilde D_2$ is defined as
\be
\tilde D_2=D_2\left(\frac{24\pi^2}{11} \right)^{2},
\ee
and we have neglected the UV subleading terms, for example $G_0^{(1)}(\phi)$ and $G_0^{(2)}(\phi)$. We have fixed the induced metric to be flat as well, and introduced the energy scale $\mu$ as the scale factor of the induced metric $\gamma_{\mu\nu}=e^{2A}\eta_{\m\n}=\big(\mu\ell\big)^2 \eta_{\mu\nu}$. The constants $D_0$ and $\tilde D_2$ are dimensionless.

The vev of ${\cal G} =Tr F^2$ is given by the functional derivative of $S^{(ren)}$ with respect to $J=\lambda ^{-1}$. At the zero-derivative order,
\begin{eqnarray}
\<\cal G\>_\lambda&=&\<Tr F^2\>=\frac {\delta S^{(ren)}}{\delta (-2\lambda)^{-1}}=\frac {8}{b_0} \big(M\ell\big)^3 \big(\mu e^{ -\frac 1 {b_0\lambda}- \A^\ast}\lambda^{-a}\big)^4\big[1+ O(\lambda)\big]\label{vev-app},
\end{eqnarray}
We have chosen the scheme with $D_0=1$ and this choice is explained in the discussion of RG invariant operator \eqref{rginv5}.

Inverting the relation between the vev $\cal G$ and the coupling $J=(-4 \lambda)^{-1}$, we can write the coupling as a function of the vev $\cal G$ and the RG scale $\mu$
\begin{align}
-\frac 1 {2\lambda}=&\frac {b_0} {8} \ln\left\{ \frac {b_0}{8} \big(M\ell\big)^{-3}\mu^{-4}e^{4\A^\ast}\lambda^{4a}	{\cal G}\big[1+O(\lambda)\big]\right\}\nonumber\\
=&\frac {b_0} {8} \ln\left[ \frac {b_0}{8} \big(M\ell\big)^{-3}\frac{\cal G}{\mu^4}\right]
+\frac {b_0}2  A^\ast
-a \,\frac {b_0}2 \ln\left[\frac  {b_0}4\ln\Big(\frac {8}{b_0}M^3\ell^3e^{-4\A^\ast}	\,\frac  {\mu^4} {\cal G}\Big)	 \right]\nonumber\\
&+O  \Big[\Big(\ln^{-1} \frac {\mu^4} {\cal G} \Big)\ln\Big(\ln \frac {\mu^4}{\cal G}\Big) \Big].
\end{align}

We then compute the effective action by Legendre transforming $S^{(ren)}$ with respect to the fluctuations of the coupling
\be
\tilde J(x)=J(x)-\bar J(\mu)=\frac 1 {2\bar\lambda(\mu)}-\frac 1 {2\lambda(x)}
\ee
around the RG scale dependent background coupling $\bar J(\mu)$. To two-derivative order the effective action \refeq{ihqcdaction-UV} reads:
\begin{align}
&\Gamma[{\cal G},\mu,\bar\lambda(\mu,\Lambda)]\nonumber \\
=&\int {\cal G} \tilde J-S^{(ren)}\nonumber\\
=&\int {\cal G} \Big\{\frac {b_0} {8} \ln\left[ \frac {b_0}{8} \big(M\ell\big)^{-3}\frac{\cal G}{\mu^4}\right]
+\frac {b_0}2  A^\ast
-a \,\frac {b_0}2 \ln\left[\frac  {b_0}4\ln\Big(\frac {8}{b_0}M^3\ell^3e^{-4\A^\ast}	\,\frac  {\mu^4} {\cal G}\Big)	 \right]\nonumber\\
&\qquad\qquad+\frac 1{2\bar\lambda} -\frac  {b_0} {8} +O  \Big[\Big(\ln^{-1} \frac {\mu^4} {\cal G} \Big)\ln\Big(\ln \frac {\mu^4}{\cal G}\Big)\Big]\Big\}\nonumber\\
&+\int\frac {b_0^2}{16}\Big[(M\ell)^3\frac {b_0}{8}{\cal G}\Big]^{\frac 1 2} {\cal G}^{-2}\frac 1 2 (\partial{\cal G})^2\left[1+O \Big(\ln^{-1} \frac {\mu^4} {\cal G} \Big)\right].
\end{align}
where for simplicity we have used the scheme with $ D_2=\left(\frac{24\pi^2}{11} \right)^{-2}$ and thus $\tilde D_2=1$.

Since $\bar J(\mu)$ is the coupling constant evaluated at energy scale $\mu$, 
one can relate the background coupling constant to the non-perturbative scale $\Lambda$
\be
\Lambda= N_c^{-\frac 1 2}(4 M^3 \ell^3)^{\frac 1 4}\mu \,e^{-\A(\bar\lambda)}
\ee
 defined in \eqref{np4} which is in one-to-one corresponce to our choice of solution.

Together with the UV expansion of $\A(\lambda)$, one can express the background coupling constant in terms of the ratio $ \mu / \Lambda$
\begin{align}\label{bg-l-UV}
\frac 1 {2\bar\lambda}=&\frac{b_0}{8}\ln\Big[4N_c^{-2}(M\ell)^{ 3}\frac {\m^4}{\Lambda^4}\Big]-\frac{b_0}2\A^\ast-a\, \frac{b_0}2\ln\bar\lambda+O(\bar\lambda)\nonumber\\
=&\frac{b_0}{8}\ln\Big[4N_c^{-2}(M\ell)^{ 3}\frac {\m^4}{\Lambda^4}\Big]-\frac {b_0}2\A^\ast-a\,\frac {b_0}2\ln\Big\{\frac 4 {b_0}\ln^{-1}\Big[4N_c^{-2}(M\ell)^{ 3}\frac {\m^4}{\Lambda^4}\Big]	\Big\}\nonumber\\
&+O  \Big[\Big(\ln^{-1} \frac {\mu} {\Lambda} \Big)\ln\Big(\ln \frac {\mu}{\Lambda}\Big)\Big].
\end{align}

Substituting the background coupling constant $\bar \lambda(\Lambda/\mu)$ by its approximated form \eqref{bg-l-UV} in the UV region, the effective action of the vev $\<{\cal G}\>$ becomes

\be
\Gamma[{\cal G},\Lambda] =\int \left[ \frac {b_0  {\cal G}}{8} \left( \ln\frac{{b_0 \cal G}}{2N_c^2\Lambda^4}  -1 \right)
+\frac {b_0^{\frac 5 2}}{32\sqrt 2}(M\ell)^{\frac 3 2} {\cal G}^{-\frac 3 2}\frac 1 2 (\partial{\cal G})^2\right],
\ee
where we have neglected the subleading terms in the $\mu\rightarrow \infty$ limit. Notice that the explicit $\mu$ dependence is cancelled at the leading order of UV expansion in both the potential and the kinetic term, so the UV effective action is RG scale independent.

From the kinetic term, the canonically normalized operator $\cO$ is determined to be:
\begin{equation}
{\cal G}^{(c)}=\frac {1} {2^{\frac 3 4}}{b_0}^{\frac {5} {4}}(M\ell)^{\frac 3 4} {\cal G}^{\frac {1}{4}}.
\end{equation}

From dimensional analysis, one can see the gluon operators have the correct scaling dimensions: ${\cal G}\sim \mu^4$ and ${\cal G}^{(c)}\sim \mu$.

In terms of the canonically normalized operator ${\cal G}^{(c)}$, the 1PI effective action of the vev $\<Tr F^2\>$ in the UV reads:
\be
\Gamma[{\cal G}^{(c)}]=\int \left\{ b_0^{-4}(M\ell)^{-3}{{{\cal G}_{(c)}}^4}\Big[ \ln \Big(\frac {{{\cal G}_{(c)}}^4}{N_c^2\Lambda^4}\Big) + \ln\Big(\frac {4}{b_0^4M^3\ell^3}\Big)-1\Big]
+ \frac 1 2\eta^{\mu\nu}\partial_\mu {{\cal G}^{(c)}}\partial_\nu {{\cal G}^{(c)}} \right\}.
\ee
This expression for the one-loop effective action realizes the field theory expectation based on the conformal anomaly (see e.g. \cite{preskill}).

\subsection{1PI effective action in the IR}
Using the IR solution of the superpotential (\ref{W-IR}), the function $\cal A(\phi$) in the renormalized generating functional action \eqref{running}  can be written as a function of $\lambda$
\begin{align}
\A(\lambda)=&-\frac 1 {6}\int^\phi d \tilde \phi \frac W {W'} =-\frac 4 9 \int \frac{d\lambda}{\lambda^2}\Big(\frac Q \lambda +\frac P 2 \frac 1 {\lambda\ln\lambda}\Big)^{-1}\nonumber\\
=&-\frac 4 {9Q}\ln \lambda+\frac {2P}{9Q^2}\ln\ln\lambda+\A^\ast+O(\ln^{-1}\lambda),
\end{align}
where the subleading terms in the $\lambda\rightarrow \infty$ limit will be neglected in the following discussion.

Using the critical value $Q=\frac 2 3$, the expotential of $\A$ is simplified into
\be\label{A-IR}
e^\A=\lambda^{-Q}\left(\ln \lambda\right)^{\frac P 2}.
\ee

Substituting $e^\A$ by its IR asymptotic form \eqref{A-IR}, the renormalized generating functional (\ref{running}) becomes:
\begin{align}
S^{(ren)}=&M^{3}\ell^{-1}\int d^4 x\,\sqrt{-\gamma}\left(D_0 e^{-4\A}+D_3\ell^{2}Q^2\frac 3 8 e^{-2\A}  \frac 1 2 \gamma^{\mu\nu}\partial_\mu\phi\partial_\nu\phi\right)\nonumber\\
=&M^{3}\ell^{3}\int d^4 x\,\left[\mu^4\lambda^{4Q}\left(\ln \lambda\right)^{-2 P}+D_3\mu^2\left(\ln \lambda\right)^{-P }\frac 1 2 \eta^{\mu\nu}\partial_\mu (\lambda^Q)\partial_\nu(\lambda^Q)\right],
\end{align}
where we have chosen the scheme with $D_0=1$ and $ D_3$ being a positive constant defined as
\begin{equation}
 D_3=\Big(D_0 \ell^{2} G_0^{(1)}(\phi_{IR})+D_1\Big)\frac  {16} {3Q^2}.
\end{equation}
The function $G_0^{(1)}(\phi)$ and $G_0^{(2)}(\phi)$ used in the calculation are defined in \eqref{G1} and \eqref{G2}. The subscript $IR$ indicates the function $G_0^{(1)}(\phi_{IR})$ is evaluated in the IR.
We have used the asymptotic form $G_0^{(2)}\rightarrow 2G_0^{(1)}W'^2$ for large $\phi$ to simplify the coefficient of the kinetic term.
The IR subleading terms have been neglected. One can see that the contribution from $G_0^{(1)}$ is important although its UV contribution is negligible.
Similar to the UV computations, we have introduced the energy scale $\mu$ as the scale factor of the flat induced metric $\gamma_{\mu\nu}=\big(\mu\ell\big)^2 \eta_{\mu\nu}$.

At the zero-derivative order, the vev $\<{\cal G}\>$ is derived from the functional derivative of the renormalized generating functional \eqref{running} with respect to $J=(-4\lambda )^{-1}$:
\begin{eqnarray}
\<{\cal G}\>_\lambda=\<Tr F^2\>=\frac {\delta S^{(ren)}}{\delta (-2\lambda)^{-1}}=8Q \big(M\ell\big)^3 \mu^4\lambda^{4Q+1}\left(\ln \lambda\right)^{-2 P}.
\end{eqnarray}

We can express the coupling as a function of the vev by inverting the relation between $\cal G$ and $J=(-2\lambda)^{-1}$
\be
-\frac 1 {2\lambda}=-\frac 1 2\left\{(8Q)^{-1} (M\ell)^{-3}\left(\frac {\cal G} {\mu^{4}}\right)\left[(4Q+1)^{-1}\ln \left(\frac {\cal G} {\mu^{4}}\right)\right]^{2P}\right\}^{-\frac 1 {4Q+1}}.
\ee

We can now calculate the effective action by Legendre transforming $S^{(ren)}$ with respect to the fluctuations of the coupling
\be
\tilde J(x)=J(x)-\bar J(\mu)=\frac 1 {2\bar\lambda(\mu)}-\frac 1 {2\lambda(x)}.
\ee
The effective action up to two-derivative order is computed using \refeq{ihqcdaction-UV}
\begin{align}
&\Gamma[{\cal G},\mu,\bar\lambda(\mu,\Lambda)]\nonumber\\
=&\int {\cal G} \tilde J-S^{(ren)}\nonumber\\
=&\int {\cal G}\Big\{\frac 1 {2\bar\lambda}-\frac 1 2 \big[1+(4Q)^{-1}\big](8Q M^3\ell^3)^{\frac 1 {4Q+1}}\Big(\frac {\cal G} {\mu^{4}}\Big)^{-\frac 1 {4Q+1}}\Big[(4Q+1)^{-1}\ln \Big(\frac {\cal G} {\mu^{4}}\Big)\Big]^{{-\frac {2P} {4Q+1}}}	\Big\}\nonumber\\
&+\int D_3\mu^2\big(8Q\big)^{-\frac {2Q}{4Q+1}}
(M^3\ell^3)^{\frac {2Q+1}{4Q+1}}
(4Q+1)^{\frac P{4Q+1}}\frac 1 2\left(\partial\Big[\Big(\frac {\cal G}{\mu^4}\Big)^{\frac Q{4Q+1}}\Big(\ln\frac {\cal G}{\mu^4}	\Big)^{-\frac P{2(4Q+1)}}	\Big]\right)^2\nonumber\\
=&\int \left\{ \frac {\cal G}  {2\bar\lambda}+D_3\mu^2\big(8Q\big)^{-\frac {2Q}{4Q+1}}
(M^3\ell^3)^{\frac {2Q+1}{4Q+1}}(4Q+1)^{\frac P{4Q+1}}\frac 1 2\left(\partial\Big[\Big(\frac {\cal G}{\mu^4}\Big)^{\frac Q{4Q+1}}\Big(\ln\frac {\cal G}{\mu^4}	\Big)^{-\frac P{2(4Q+1)}}	\Big]\right)^2\right\}
\end{align}
where in the last line we have neglected the subleading terms in the limit $\cal G/\mu\rightarrow \infty$ in the potential and we can see that the effective potential has a linear dependence on the vev $\cal G$ with a slope $(4\bar\lambda)^{-1}$.

Then we need to express the background coupling constant in terms of the non-perturbative scale \eqref{np6}
\be
\Lambda=N_c^{-\frac 1 2} (4 M^3 \ell^3)^{\frac 1 4}\mu \,e^{-\A(\bar\lambda)},
\ee
the same as what have been done in the the UV expansion. Inverting the relation between $\Lambda$ and $\bar\lambda$, the background coupling as a function of the RG scale $\mu$ is
\be\label{bg-l-IR}
\frac 1 {\bar\lambda}=\left[(4 M^3\ell^3)^{-\frac 1 4}N_c^{\frac 1 2}\Big(\frac \Lambda \mu \Big)\ln^{\frac P 2}\bar\lambda\right]^{-\frac 1 Q}
=(4M^3\ell^3)^{\frac 1 {4Q}}Q^{\frac P{2Q}}\Big(\frac {N_c^{\frac 1 2}\Lambda} \mu\Big)^{-\frac 1 Q}\ln^{-\frac P{2Q}}\Big(\frac {N_c^{\frac 1 2}\Lambda} \mu\Big),
\ee
where we have kept only the IR leading terms and neglected subleading terms in the $\mu/\Lambda\rightarrow 0$ limit.

Substituting the background coupling $\bar J=(-4\bar \lambda)^{-1}$  by its IR asymptotics \eqref{bg-l-IR}, we have
\begin{align}
\Gamma[{\cal G},\mu]=&\int \frac 1 2(4M^3\ell^3)^{\frac 1 {4Q}}Q^{\frac P{2Q}}\Big(\frac {N_c^{\frac 1 2}\Lambda} \mu\Big)^{-\frac 1 Q}\ln^{-\frac P{2Q}}\Big(\frac {N_c^{\frac 1 2}\Lambda} \mu\Big)\,{\cal G}\nonumber\\
&+\int D_3\mu^2\big(8Q\big)^{-\frac {2Q}{4Q+1}}
(M^3\ell^3)^{\frac {2Q+1}{4Q+1}}(4Q+1)^{\frac P{4Q+1}}\frac 1 2\partial\Big[\Big(\frac {\cal G}{\mu^4}\Big)^{\frac Q{4Q+1}}\Big(\ln\frac {\cal G}{\mu^4}	\Big)^{-\frac P{2(4Q+1)}}	\Big]^2	
\end{align}

From the kinetic term, one determines the canonically normalized operator as
\be
{\cal G}^{(c)}=D_3^{\frac 1 2} \mu \big(8Q\big)^{-\frac {Q}{4Q+1}}
(M^3\ell^3)^{\frac {2Q+1}{2(4Q+1)}}(4Q+1)^{\frac P{2(4Q+1)}}\Big(\frac {\cal G}{\mu^4}\Big)^{\frac Q{4Q+1}}\Big(\ln\frac {\cal G}{\mu^4}	\Big)^{-\frac P{2(4Q+1)}}.
\ee

Then we can invert the relation between ${\cal G}^{(c)}$ and $\cal G$
\begin{align}
{\cal G}=&8Q(M^3\ell^3)^{-\frac {2Q+1} {2Q}} (4Q+1)^{-\frac P {2Q}}\mu^4 \left(\frac {{\cal G}^{(c)}}{D_3^{\frac 1 2}\mu}\right)^{\frac {4Q+1}Q}\left(\frac {4Q+1}Q\right)^{\frac P {2Q}}\ln ^{\frac P{2Q}}\left(\frac {{\cal G}^{(c)}}{D_3^{\frac 1 2}\mu}\right)\nonumber\\
=&8(M^3\ell^3)^{-\frac {2Q+1} {2Q}}Q^{1-\frac P{2Q}} \mu^4 \left(\frac {{\cal G}^{(c)}}{D_3^{\frac 1 2}\mu}\right)^{\frac {4Q+1}Q}\ln ^{\frac P{2Q}}\left(\frac {{\cal G}^{(c)}}{\mu}\right)
\end{align}

Finally, in terms of the canonically normalized operator ${\cal G}^{(c)}$, the effective action reads:
\begin{align}
\Gamma[{\cal G},\mu,\bar\lambda(\mu,\Lambda)]=\int\left[ D_4\left(\ln\frac {N_c^{\frac 1 2}\Lambda} \mu\right)^{-\frac P{2Q}} {{\cal G}_{(c)}}^{4}\left(\frac {{\cal G}^{(c)}}{N_c^{\frac 1 2}\Lambda}\right)^{\frac 1 Q}\left(\ln \frac {{\cal G}^{(c)}}{\mu}\right)^{\frac P{2Q}}+ \frac 1 2\eta^{\mu\nu}\partial_\mu{\cal G}^{(c)}\partial_\nu{\cal G}^{(c)}\right],
\end{align}
\be
D_4=(4M^{-3}\ell^{-3})^{\frac {4Q+1} {4Q}}Q\left[\Big(\ell^{2} G_0^{(1)}(\phi_{IR})+D_1\Big)\frac  {16} {3Q^2}\right]^{-\frac {4Q+1} {2Q}},
\ee
where $G_0^{(1)}(\phi)$ is defined in \eqref{G1} and the subscript $IR$ indicates it is evaluated in the IR.



\begin{thebibliography}{100}

\def\hri#1#2{\href{http://arxiv.org/abs/#1}{[ArXiv:#1]#2}}
\def\hre#1#2{\href{http://arxiv.org/abs/#1/#2}{[ArXiv:#1/#2]}}
\def\hspi#1#2{\href{http://www.slac.stanford.edu/spires/find/hep/www?irn=#1}{#2}}

\bibitem{gkn}
  U.~Gursoy and E.~Kiritsis,
  {\em ``Exploring improved holographic theories for QCD: Part I,''}
  JHEP {\bf 0802} (2008) 032
  \hri{0707.1324}{[hep-th]};\\
  U.~Gursoy, E.~Kiritsis and F.~Nitti,
  {\em ``Exploring improved holographic theories for QCD: Part II,''}
  JHEP {\bf 0802} (2008) 019
  \hre{0707.1349}{[hep-th]}.
  
  \bibitem{gubser}
  S.~S.~Gubser and A.~Nellore,
  {\em ``Mimicking the QCD equation of state with a dual black hole,''}
  Phys.\ Rev.\ D {\bf 78} (2008) 086007
  \hri{0804.0434}{[hep-th]}.

  
\bibitem{thermo2}
U.~Gursoy, E.~Kiritsis, L.~Mazzanti and F.~Nitti,
{\em ``Holography and Thermodynamics of 5D Dilaton-gravity,''}
  JHEP {\bf 0905} (2009) 033
  \hri{0812.0792}{[hep-th]}.

\bibitem{thermo3}
U.~Gursoy, E.~Kiritsis, L.~Mazzanti and F.~Nitti,
  {\em``Improved Holographic Yang-Mills at Finite Temperature: Comparison with Data,''}
  Nucl.\ Phys.\ B {\bf 820}, 148 (2009)
  \hri{0903.2859}{[hep-th]}.

\bibitem{diss}
E.~Kiritsis,
  {\em ``Dissecting the string theory dual of QCD,''}
  Fortsch.\ Phys.\  {\bf 57}, 396 (2009)
  \hri{0901.1772}{[hep-th]}.

\bibitem{ihqcd}
  U.~Gursoy, E.~Kiritsis, L.~Mazzanti, G.~Michalogiorgakis and F.~Nitti,
  {\em ``Improved Holographic QCD,''}
  Lect.\ Notes Phys.\  {\bf 828}, 79 (2011)
  \hri{1006.5461}{[hep-th]}.

\bibitem{kln}
E.~Kiritsis, W.~Li and F.~Nitti,
  {\em ``Holographic RG flow and the Quantum Effective Action,''}
 Fortschr.\ Phys. {\bf  62}, No. 5-6, 389-454 (2014)
 [arXiv:1401.0888 [hep-th]]



\bibitem{svz}
M.~A.~Shifman, A.~I.~Vainshtein and V.~I.~Zakharov,
  {\em ``QCD and Resonance Physics. Sum Rules,''}
  Nucl.\ Phys.\ B {\bf 147}, 385 (1979).



\bibitem{lattice}
A.~Di Giacomo and G.~C.~Rossi,
  {\em ``Extracting the Vacuum Expectation Value of the Quantity alpha / pi G G from Gauge Theories on a Lattice,''}
  Phys.\ Lett.\ B {\bf 100}, 481 (1981).\\
T.~Banks, R.~Horsley, H.~R.~Rubinstein and U.~Wolff,
  {\em ``Estimate of the Gluon Condensate From Monte Carlo Calculations,''}
  Nucl.\ Phys.\ B {\bf 190}, 692 (1981).\\
  M.~Campostrini, A.~Di Giacomo and Y.~Gunduc,
  Phys.\ Lett.\ B {\bf 225}, 393 (1989).

\bibitem{latticeT}
G.~Boyd and D.~E.~Miller,
  {\em ``The Temperature dependence of the SU(N(c)) gluon condensate from lattice gauge theory,''}
  hep-ph/9608482.\\
D.~E.~Miller,
  Acta Phys.\ Polon.\ B {\bf 28}, 2937 (1997)
  [hep-ph/9807304].






\bibitem{baumann}
D.~Baumann and D.~Green,
  {\em ``Desensitizing Inflation from the Planck Scale,''}
  JHEP {\bf 1009}, 057 (2010)
  [arXiv:1004.3801 [hep-th]].

\bibitem{piai}
D.~Elander, A.~F.~Faedo, C.~Hoyos, D.~Mateos and M.~Piai,
  {\em ``Multiscale confining dynamics from holographic RG flows,''}
  JHEP {\bf 1405}, 003 (2014)
  [arXiv:1312.7160 [hep-th], arXiv:1312.7160].


\bibitem{megias}
  E.~Megias and O.~Pujolas,
  {\em ``Naturally light dilatons from nearly marginal deformations,''}
  arXiv:1401.4998 [hep-th].











\bibitem{gubser1} 
  S.~S.~Gubser,
 {\em ``Curvature singularities: The Good, the bad, and the naked,''}
  Adv.\ Theor.\ Math.\ Phys.\  {\bf 4}, 679 (2000)
   \hri{ [hep-th/0002160]}.
  
   \bibitem{5dgraviton}
 E.~Kiritsis and F.~Nitti,
   {\em ``On massless 4D gravitons from asymptotically AdS(5) space-times,''}
   Nucl.\ Phys.\ B {\bf 772}, 67 (2007)
   \hre{hep-th}{0611344}.
   
   
   \bibitem{papa}
  I.~Papadimitriou,
  {\em ``Holographic Renormalization of general dilaton-axion gravity,''}
  JHEP {\bf 1108} (2011) 119
\hri{1106.4826}{[hep-th]}.

 \bibitem{bianchi}
   M.~Bianchi, D.~Z.~Freedman and K.~Skenderis,
   {\em ``How to go with an RG flow,''}
   JHEP {\bf 0108}, 041 (2001)
   \hre{hep-th}{0105276};\\
 M.~Bianchi, D.~Z.~Freedman and K.~Skenderis,
   {\em ``Holographic renormalization,''}
   Nucl.\ Phys.\ B {\bf 631}, 159 (2002)
   \hre{hep-th}{0112119}.

 



\bibitem{KN}
  E.~Kiritsis and V.~Niarchos,
  {\em ``The holographic quantum effective potential at finite temperature and density,''}
  JHEP {\bf 1208} (2012) 164
  \hri{1205.6205}{[hep-th]}.

\bibitem{preskill}
J.~Schechter,
  {\em ``Effective Lagrangian with Two Color Singlet Gluon Fields,''}
  Phys.\ Rev.\ D {\bf 21}, 3393 (1980).


\bibitem{chenetal}
C.~J.~Morningstar and M.~J.~Peardon,
  {\em ``The glueball spectrum from an anisotropic lattice study,''}
  Phys.\ Rev.\  D {\bf 60}, 034509 (1999)
  \hre{hep-lat}{9901004};\\
 Y.~Chen {\it et al.},
  {\em ``Glueball spectrum and matrix elements on anisotropic lattices,''}
  Phys.\ Rev.\  D {\bf 73}, 014516 (2006)
  \hre{hep-lat}{0510074}.

\end{thebibliography}
\end{document}